\documentclass[12pt]{article}

\usepackage{graphics,epsfig}

\newcommand{\bea}{\begin{eqnarray}}
\newcommand{\eea}{\end{eqnarray}}
\newcommand{\beq}{\begin{equation}}
\newcommand{\eeq}{\end{equation}}
\newcommand{\nn}{\nonumber}
\newcommand{\gev}{{\rm GeV}}
\newcommand{\mev}{{\rm MeV}}
\newcommand{\fm}{{\rm fm}}
\newcommand{\msb}{\overline{\rm{MS}}}

\def\dfrac#1#2{{\displaystyle {#1 \over #2}}}

\def\simge{\mathrel{\rlap{\raise 0.511ex \hbox{$>$}}{\lower 0.511ex
 \hbox{$\sim$}}}}
\def\simle{\mathrel{\rlap{\raise 0.511ex \hbox{$<$}}{\lower 0.511ex
 \hbox{$\sim$}}}}
\def\slash#1{\setbox0=\hbox{$#1$}\dimen0=\wd0 \setbox1=\hbox{/} \dimen1=\wd1
 \ifdim\dimen0>\dimen1 \rlap{\hbox to \dimen0{\hfil/\hfil}} #1
 \else \rlap{\hbox to \dimen1{\hfil$#1$\hfil}} / \fi}

\def\desy{a}
\def\ors{b}
\def\rmii{c}
\def\rmiii{d}
\def\infn{e}
\def\barc{f}
\def\val{g}
\def\liv{h}
\def\hum{i}
\def\ber{j}

\begin{document}
\begin{titlepage}
{
\normalsize
\hfill \parbox{100mm}{DESY 09-044,~FTUV-09-0112,~ICCUB-09-193,\\
                      IFIC/09-02,~LPT-Orsay/09-18,~LTH-828,\\
                      ROM2F/2009/05,~RM3-TH/09-7,~SFB/CPP-09-32,\\
                      UB-ECM-PF~09/08}}\\[10mm]
\begin{center}
  \begin{Large}
    \textbf{Pseudoscalar decay constants of kaon and $D$-mesons \\
            from $N_f=2$ twisted mass Lattice QCD\unboldmath} \\
  \end{Large}
\end{center}

\begin{figure}[h]
  \begin{center}
    \includegraphics[draft=false]{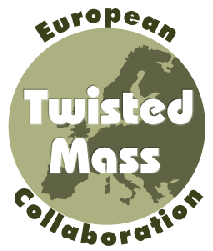}
  \end{center}
\end{figure}

\baselineskip 20pt plus 2pt minus 2pt
\begin{center}
  \textbf{
    B.~Blossier$^{(\desy,\ors)}$,
    P.~Dimopoulos$^{(\rmii)}$,
    R.~Frezzotti$^{(\rmii)}$,
    B.~Haas$^{(\ors)}$,\\
    G.~Herdoiza$^{(\desy)}$,
    K.~Jansen$^{(\desy)}$,
    V.~Lubicz$^{(\rmiii,\infn)}$,
    F.~Mescia$^{(\barc)}$,
    D.~Palao$^{(\val)}$,\\
    A.~Shindler$^{(\liv)}$,
    S.~Simula$^{(\infn)}$,
    C.~Tarantino$^{(\rmiii,\infn)}$,
    C.~Urbach$^{(\hum)}$,
    U.~Wenger$^{(\ber)}$}\\
\end{center}

\begin{center}
  \begin{footnotesize}
    \noindent

$^{(\desy)}$ DESY, Zeuthen, Platanenallee 6, D-15738 Zeuthen, Germany
\vspace{0.2cm}

$^{(\ors)}$ Laboratoire de Physique Th\'eorique (B\^at.~210), Universit\'e de
Paris XI,\\ Centre d'Orsay, 91405 Orsay-Cedex, France
\vspace{0.2cm}

$^{(\rmii)}$ Dip. di Fisica, Universit{\`a} di Roma Tor Vergata and INFN, Sez.
di Roma Tor Vergata,\\ Via della Ricerca Scientifica, I-00133 Roma, Italy
\vspace{0.2cm}

$^{(\rmiii)}$ Dip. di Fisica, Universit{\`a} di Roma Tre, Via della Vasca Navale
84, I-00146 Roma, Italy
\vspace{0.2cm}

$^{(\infn)}$ INFN, Sez. di Roma
III, Via della Vasca Navale 84, I-00146 Roma, Italy
\vspace{0.2cm}

$^{(\barc)}$ Departament d'Estructura i Constituents de la Mat\`eria and
Institut de Ci\`encies del Cosmos (ICCUB),
Universitat de Barcelona, Diagonal 647, 08028 Barcelona, Spain
\vspace{0.2cm}

$^{(\val)}$ Dep. de F\'{\i}sica Te\`{o}rica and IFIC, Univ. de Val\`{e}ncia,
Dr.Moliner 50, E-46100 Burjassot, Spain
\vspace{0.2cm}

$^{(\liv)}$ Theoretical Physics Division, Dept. of Mathematical Sciences,
\\University of Liverpool, Liverpool L69 7ZL, UK
\vspace{0.2cm}

$^{(\hum)}$ Institut f\"ur Elementarteilchenphysik, Fachbereich Physik,
\\ Humbolt Universit\"at zu Berlin, D-12489, Berlin, Germany
\vspace{0.2cm}

$^{(\ber)}$ Institute for Theoretical Physics, University of Bern, Sidlerstr.
5, CH-3012 Bern, Switzerland
\vspace{0.2cm}

  \end{footnotesize}
\end{center}
\end{titlepage}

\begin{abstract}{
We present the results of a lattice QCD calculation of the pseudoscalar meson
decay constants  $f_\pi$, $f_K$, $f_D$ and $f_{D_s}$, performed  with $N_f=2$
dynamical fermions. The simulation is carried out with the tree-level improved
Symanzik gauge action and with the twisted mass fermionic action at maximal
twist. We have considered for the final analysis three values of the lattice
spacing, $a\simeq 0.10$ fm, 0.09 fm and 0.07 fm, with pion masses down to $m_\pi
\simeq 270\ \mev$. Our results for the light meson decay constants are $f_K=
158.1(2.4)\ \mev$ and $f_K/f_\pi=1.210(18)$. From the latter ratio, by using the
experimental determination of $\Gamma(K \to \mu \bar\nu_\mu(\gamma))/
\Gamma(\pi\to \mu \bar \nu_\mu (\gamma))$ and the average value of $\vert
V_{ud}\vert$ from nuclear beta decays, we obtain $\vert V_{us}\vert=0.2222(34)$,
in good agreement with the determination from semileptonic $K_{l3}$ decays and
the unitarity constraint. For the $D$ and $D_s$ meson decay constants we obtain
$f_D=197(9)\ \mev$, $f_{D_s}=244(8)\ \mev$ and $f_{D_s}/f_D=1.24(3)$. Our result
for $f_D$ is in good agreement with the CLEO experimental measurement. For
$f_{D_s}$ our determination is smaller than the PDG 2008 experimental average
but in agreement with a recent improved measurement by CLEO at the $1.4\,\sigma$
level.}
\end{abstract}


\section{Introduction}
\label{sec:intro}
An accurate lattice determination of the pseudoscalar decay constants of kaon
and $D$-mesons is an important task. On the one hand, the ratio $f_K/f_\pi$ can
be used together with the experimental determination of $\Gamma(K \to \mu \bar
\nu_\mu (\gamma))/\Gamma(\pi \to \mu \bar \nu_\mu (\gamma))$ and the average
value of $\vert V_{ud}\vert$ from nuclear beta decays to achieve a precise
determination of the CKM matrix element $\vert V_{us}
\vert$~\cite{Marciano:2004uf} and to test the CKM first raw unitarity relation.
On the other hand, the pseudoscalar decay constants $f_D$ and $f_{D_s}$ have
been recently measured at CLEO, BaBar and Belle~\cite{PDG,Alexander:2009ux},
asking for accurate lattice determinations to be compared to.

In this paper, we present a lattice QCD calculation of the pseudoscalar meson
decay constants $f_\pi$, $f_K$, $f_D$ and $f_{D_s}$. With respect to our
previous study of the pion and kaon decay constants~\cite{Blossier:2007vv}, here
we have analysed data at three values of the lattice spacing, $a\simeq 0.10\,
\fm, 0.09\, \fm, 0.07\, \fm$ (corresponding to $\beta=3.8, 3.9, 4.05$), and
performed a chiral extrapolation taking lattice artefacts into account.
Estimating the lattice artefacts turns out to be crucial for an accurate
determination of $f_D$ and $f_{D_s}$ since cutoff effects induced by the charm
mass, which are parametrically of ${\cal O}(a^2\, m_c^2)$, are not small in our
simulation, $\sim 5 \div 10\%$. Both SU(2) and SU(3) chiral perturbation theory
(ChPT) has been considered for the chiral extrapolation, whereas only the latter
was considered in~\cite{Blossier:2007vv}. With respect to
ref.~\cite{Blossier:2007vv}, we have also added ensembles with a lighter quark
mass ($m_\pi \simeq 270\,\mev$) and a larger volume ($L \simeq 2.7\, \fm$), both
at the value of $a \simeq 0.09\,\fm$.

The calculation is based on the gauge field configurations generated by the
European Twisted Mass Collaboration (ETMC) with the tree-level improved Symanzik
gauge action~\cite{Weisz82}  and the twisted mass action~\cite{FGSW01}  at
maximal twist, discussed in detail in
refs.~\cite{Boucaud:2007uk}-\cite{Dimopoulos:2008sy}. We simulated $N_f=2$
dynamical quarks, taken to be degenerate in mass, whose masses are eventually
extrapolated to the physical isospin averaged mass of the up and down quarks.
The strange and charm quarks are quenched in the present calculation.

The use of the twisted mass fermions turns out to be beneficial, since the
pseudoscalar meson masses and decay constants, which represent the basic
ingredients of the calculation, are automatically improved at ${\cal
O}(a)$~\cite{Frezzotti:2003ni} (see also~\cite{Shindler:2007vp}), and the
determination of the pseudoscalar decay constants does not require the
introduction of any renormalization constant. Both these features allow to
significantly improve the accuracy of the calculation. It has also been shown
that, for twisted mass fermions at maximal twist, the so called KLM
factor~\cite{ElKhadra:1996mp}, which relates the lattice quark propagator at
zero momentum to the continuum one, is equal to one at tree level, to all order
in $a m_q$~\cite{McNeile:2008wr}. This is beneficial in reducing discretization
effects particularly for heavy quark masses, as the charm quark mass considered
in this study.

As discussed in refs.~\cite{Blossier:2007vv,Frezzotti:2004wz,AbdelRehim:2006ve},
we implement non-degenerate valence quarks in the twisted mass formulation by
formally introducing a twisted doublet for each non-degenerate quark flavour. In
the present analysis we thus introduce in the valence sector three twisted
doublets, ($u,d$), ($s,s^\prime$) and ($c,c^\prime$), with masses $\mu_l$,
$\mu_s$ and $\mu_c$ respectively. Within each doublet, the two valence quarks
are regularized in the physical basis with Wilson parameters of opposite values
($r=-r^\prime=1$). We simulate mesons composed of quarks with opposite Wilson
parameters so that the squared meson mass $m^2_{PS}$ differs from its continuum
counterpart only by terms of ${\cal O}(a^2\,\mu_q)$ and ${\cal O}(a^4)$, whereas
$f_{PS}$ differs from its continuum limit by terms of ${\cal
O}(a^2)$~\cite{Sharpe:2004ny,Frezzotti:2005gi}. This implies that at ${\cal
O}(a^2)$ the cutoff effects on $m^2_{PS}$ and $f_{PS}$ are as in a chiral
invariant lattice formulation. In our calculation large artefacts, like those
affecting the neutral pion mass in the twisted mass formulation of lattice QCD,
seem not to be present.

In the present analysis, we study the pseudoscalar decay constants as a function
of the meson masses, whereas in our previous work~\cite{Blossier:2007vv} we
relied on their dependence on the quark masses. When data at different values of
the lattice spacing are involved, the study in terms of meson masses is simpler,
since it does not require the introduction of the quark mass renormalization
constant ($Z_m = Z^{-1}_P$) to convert at each lattice spacing from the bare to
the renormalized (cutoff independent) quark mass. The dependence of the decay
constants on the meson masses is studied together with the dependence on the
lattice spacing, through a combined fit where terms of ${\cal O}(a^2)$, coming
from the Symanzik expansion of the lattice theory, are added to the functional
forms predicted by ChPT. In this way, the continuum and chiral extrapolations of
the lattice results are performed simultaneously. In alternative to this
combined analysis, a different approach could be adopted. It consists of first
extrapolating data at fixed meson mass values to the continuum, and then
extrapolating the obtained continuum results to the physical point. With our
simulation setup, however, this procedure turns out to be unsafe, since for some
values of the meson masses we have data at only two values of the lattice
spacing. Such a procedure could become feasible when data at a smaller value of
the lattice spacing are available. Corresponding simulations with $a \simeq
0.05\, \fm$ are currently performed by ETMC.

In order to perform the extrapolation to the physical masses we have used ChPT
for the light mesons and Heavy Meson ChPT (HMChPT)~\cite{Sharpe:1995qp} for the
$D$ sector. For the kaon and $D_s$ mesons we have considered both SU(2)- and
SU(3)-ChPT. In the SU(2) case one treats the $u/d$ quarks  as light, while the
strange quark is not required to satisfy chiral symmetry. The short
interpolation to the physical strange quark, which is required in our analysis,
is then performed linearly~\cite{Allton:2008pn}. This is justified, since our
simulated values of the strange quark mass are around the physical mass. For
comparison, we have also considered chiral extrapolations based on SU(3)-ChPT
and SU(3)-HMChPT. We find that the SU(2) effective theory, which is less
predictive than SU(3), provides however a better description of the lattice data
for the decay constants up to the region of the strange quark mass.

Our final results for the kaon and $D$-mesons decay constants are given in the
abstract and in eqs.~(\ref{eq:fKfinal}) and (\ref{eq:fDfinal}). Our
determination of $f_K/f_\pi$ leads to a determination of $\vert V_{us}\vert$
that is in good agreement with the value obtained from semileptonic kaon decays,
though with a larger error, as well as with the first row unitarity constraint
of the CKM matrix. Of relevant phenomenological interest is also our result for
$f_{D_s}$, which is about $2.3\,\sigma$ lower than the experimental average
quoted by the PDG~\cite{PDG} but in agreement with other unquenched lattice
determinations and with a recent improved measurement by
CLEO~\cite{Alexander:2009ux} at the $1.4\, \sigma$ level. The value indicated by
the new CLEO measurement weakens the tension between experimental and lattice
results for $f_{D_s}$, which suggested explanations in terms of new physics
effects~\cite{Dobrescu:2008er,Akeroyd:2009tn}.

The plan of this paper is as follows. In section 2 we provide the details of the
lattice simulations used for the present study and discuss the determination of
the pseudoscalar meson masses and decay constants. The combined chiral and
continuum extrapolation fits of the light and $D$-mesons decay constants are
discussed in sections 3 and 4, respectively. There, we also provide our final
results for the decay constants, discussing in particular the evaluation of the
systematic uncertainties and the comparison with other lattice determinations
and with recent experimental measurements.

\section{Simulation details}
Details of the ensembles of gauge configurations used in the present analysis
and the values of the simulated valence quark masses are collected in
Tables~\ref{tab:simdet} and~\ref{tab:val}, respectively.
\begin{table}[t]
\begin{center}
\begin{tabular}{||c||c|c|c|c|c|c|c|c|c||} \hline 
 Ens.   &  $\beta$ &  $a\,[\fm]$ & $V/a^4$ & $a \mu_{sea}$ & $m_\pi\,[\mev]$ & 
$m_\pi \, L$ & $N_{cfg}$ & $(\Delta t)_{\pi,K}$ & $(\Delta t)_{D,D_s}$ \\
\hline\hline
$A_2$   & $3.8$ & $0.10$  &$24^3 \times 48$& $0.0080$   & $410$ & $5.0$ & $240$
&  $[10,23]$  & $[14,23]$  \\ 
$A_3$   &       &         &                & $0.0110$   & $480$ & $5.8$ & $240$
&             &            \\ 
\hline
$B_1$   & $3.9$ & $0.085$ &$24^3 \times 48$& $0.0040$   & $315$ & $3.3$ & $480$
&  $[12,23]$  & $[16,23]$  \\ 
$B_2$   &       &         &                & $0.0064$   & $390$ & $4.0$ & $240$
&             &            \\ 
$B_3$   &       &         &                & $0.0085$   & $450$ & $4.7$ & $240$
&             &            \\ 
$B_4$   &       &         &                & $0.0100$   & $490$ & $5.0$ & $240$
&             &            \\ 
\hline
$B_7$   & $3.9$ & $0.085$ &$32^3 \times 64$& $0.0030$   & $270$ & $3.7$ & $240$
&  $[12,31]$  & $[16,31]$  \\ 
$B_6$   &       &         &                & $0.0040$   & $310$ & $4.3$ & $240$
&    &   \\ 
\hline
$C_1$   &$4.05$ & $0.065$ &$32^3 \times 64$& $0.0030$   & $310$ & $3.3$ & $144$
&  $[16,31]$  & $[21,31]$  \\ 
$C_2$   &       &         &                & $0.0060$   & $430$ & $4.6$ & $128$
&             &            \\ 
$C_3$   &       &         &                & $0.0080$   & $500$ & $5.3$ & $128$
&             &            \\ 
\hline
\end{tabular}
\end{center}
\vspace{-0.4cm}
\caption{\sl Details of the ensembles of gauge configurations used in the
present study: value of the gauge coupling $\beta$; approximate value of the
lattice spacing $a$; lattice size $V=L^3 \times T$ in lattice units; bare sea
quark mass in lattice units; approximate value of the pion mass; approximate
value of the product $m_\pi \, L$; number of independent configurations
$N_{cfg}$. We also provide the fitting time interval in lattice units chosen for
the two-point pseudoscalar correlators in the pion and kaon sectors, $(\Delta
t)_{\pi,K}$, and in the $D$-meson sectors, $(\Delta t)_{D,D_s}$.}
\label{tab:simdet}
\end{table}
\begin{table}[t]
\begin{center}
\begin{tabular}{||c||c|c|c|c|c||}
\hline
  &  $A_2 - A_3$ &   $B_1 - B_4$    & $B_7$        & $B_6$        & $C_1 - C_3$
\\ \hline\hline
$a \mu_l$ 
& $0.0080$, $0.0110$ & $0.0040$, $0.0064$, $0.0085$, & $0.0030$ & $0.0040$ &
$0.0030$, $0.0060$, \\
&   & $0.0100$  &  &  & $0.0080$ \\ \hline
$a \mu_s$
& $0.020$, $0.025$, $0.030$, & $0.022$, $0.027$, $0.032$ & $0.022$, &
$0.022$, & $0.015$, $0.018$ \\
& $0.036$ &  & $0.027$ & $0.027$ & $0.022$, $0.026$ \\
\hline
$a \mu_c$
& $0.270$, $0.310$, $0.355$, & $0.250$, $0.320$, $0.390$, & $0.250$, &
$0.250$, & $0.200$, $0.230$, \\
& $0.435$, $0.520$ & $0.460$ & $0.320$ & $0.320$ &  $0.260$  $0.315$ \\
\hline
\end{tabular}
\end{center}
\vspace{-0.4cm}
\caption{\sl Values of simulated valence quark masses in lattice units for each
configuration ensemble in the light, strange and charm sectors.}
\label{tab:val}
\end{table}

Measurements are performed over independent gauge configurations that are
separated by $20$ trajectories in the case of the ensembles at $\beta=3.8$ and
$3.9$, and by $40$ trajectories in the case of the ensembles at $\beta=4.05$.
The trajectory length is equal to unity for the ensembles $A_2$, $A_3$ and $B_7$
and to $1/2$ for the ensembles $B_1$-$B_4$, $B_6$ and $C_1$-$C_3$. Among the
available ETMC ensembles we have excluded from this study those corresponding to
pion masses larger than $500\, \mev$ (ensembles $A_4$, $B_5$ and $C_4$ of
ref.~\cite{Urbach:2007rt}).

Our strategy and the conditions to tune to maximal twist have been discussed in
refs.~\cite{Boucaud:2007uk}-\cite{Urbach:2007rt}. Whereas at $\beta=3.9$ and
$\beta=4.05$ these conditions are accurately fulfilled, at $\beta=3.8$ the
situation is different. Due to large fluctuations and long autocorrelations for
the PCAC mass appearing at the smallest value of the twisted mass parameter at
$\beta=3.8$, we cannot be confident that the maximal twist condition is realized
with the same accuracy as at $\beta=3.9$ and $\beta=4.05$. In the present study,
in order to check for effects of such a possible mismatch, besides not
considering the ensemble with the lightest quark mass at $\beta=3.8$ (ensemble
$A_1$ of ref.~\cite{Urbach:2007rt}), we have also performed an analysis with and
without taking the whole set of data at $\beta=3.8$ into account. As we will
demonstrate below, fully consistent results are obtained. This finding is also
supported by the results of a theoretical analysis of the effects of being out
of maximal twist, which shows that these systematic effects on the pseudoscalar
decay constants analyzed here at $\beta=3.8$ are small compared to statistical
and other systematic uncertainties on the same data.\footnote{For the basic
ideas of this analysis in the unitary case see ref.~\cite{Dimopoulos:2007qy}.}

With respect to our previous determination of $f_\pi$ and
$f_K$~\cite{Blossier:2007vv}, the new ensembles used in the present analysis are
those with $\beta=3.8$ ($A_2$-$A_3$) and $\beta=4.05$ ($C_1$-$C_3$) and the
ensembles $B_6$-$B_7$ at $\beta=3.9$. The ensemble $B_7$ has the lightest
simulated mass $\mu_l \sim 0.15\, m_s^{phys.}$, where $m_s^{phys.}$ is the
physical strange quark mass. The ensembles $B_1$ and $B_6$ have the same value
of $\beta$ and sea quark mass but different volumes, $L\simeq 2.0\, \fm$ and
$L\simeq 2.7\, \fm$, thus allowing a study of finite size effects.

In order to investigate the properties of the $\pi$, $K$, $D$ and $D_s$ mesons,
we simulate the sea and valence light ($u/d$) quark mass in the range  $0.15\,
m_s^{phys.} \simle \mu_l \simle 0.5\, m_s^{phys.}$, the valence strange quark
mass is within $0.9\, m_s^{phys.} \simle \mu_s \simle 1.5\, m_s^{phys.}$, and
the valence charm quark mass within $0.8\, m_c^{phys.} \simle \mu_c \simle 1.5\,
m_c^{phys.}$, $m_c^{phys.}$ being the physical charm mass. The values of the
valence quark masses simulated for each configuration ensemble are collected in
Table~\ref{tab:val}. These values can be converted into the corresponding values
of the renormalized quark masses (e.g. in the $\msb$ scheme) using the available
determinations of renormalization constants given in
refs.~\cite{renPROC07,renPAPER09}.

As already noted, with twisted mass fermions at maximal twist, the determination
of the pseudoscalar decay constants, besides being automatically improved at
${\cal O}(a)$, does not require the introduction of any renormalization
constants. For a pseudoscalar meson of mass $m_{PS}$, composed of valence quarks
with masses $\mu_{val}^{(1)}$  and $\mu_{val}^{(2)}$, the decay constant
$f_{PS}$ is obtained as
\beq
f_{PS} = \left(\mu_{val}^{(1)} + \mu_{val}^{(2)}\right) \, \frac{\vert \langle 0
\vert P \vert PS \rangle\vert}{m_{PS}\,\sinh m_{PS}} \, ,
\label{eq:fPS} 
\eeq
where $P=\bar q_1 \gamma_5 q_2$. The meson mass $m_{PS}$ and the matrix element
$\vert\langle 0 \vert P \vert PS \rangle\vert$ entering eq.~(\ref{eq:fPS}) have
been extracted from a single state fit of the two-point pseudoscalar correlation
function within the time intervals collected in Table~\ref{tab:simdet}. The
replacement of $m_{PS}$ with $\sinh m_{PS}$ in the lattice
definition~(\ref{eq:fPS}) of the decay constant helps in reducing discretization
errors for heavy meson masses. Combined with the observation that the tree-level
KLM factor for the quark field is equal to one for twisted mass fermions at
maximal twist~\cite{McNeile:2008wr}, this replacement allows to remove at tree
level all ${\cal O}((a m_c)^n)$ terms in the determination of the
$D_{(s)}$-meson decay constant.

The statistical accuracy of the meson correlators is improved by using the
so-called ``one-end" stochastic method, implemented in
ref.~\cite{McNeile:2006bz} (see also~\cite{Boucaud:2008xu}), which includes all
spatial sources at a single timeslice. Statistical errors on the meson masses
and decay constants are evaluated using the jackknife procedure, with $16$
jackknife bins for each configuration ensemble. Statistical errors on the fit
results which are based on data obtained from independent ensembles of gauge
configurations are evaluated using a bootstrap procedure, with $100$ bootstrap
samples.
\begin{figure}[t]
\vspace{-1.0cm}
\includegraphics[scale=0.3,angle=270,trim=75 0 0 0, clip]{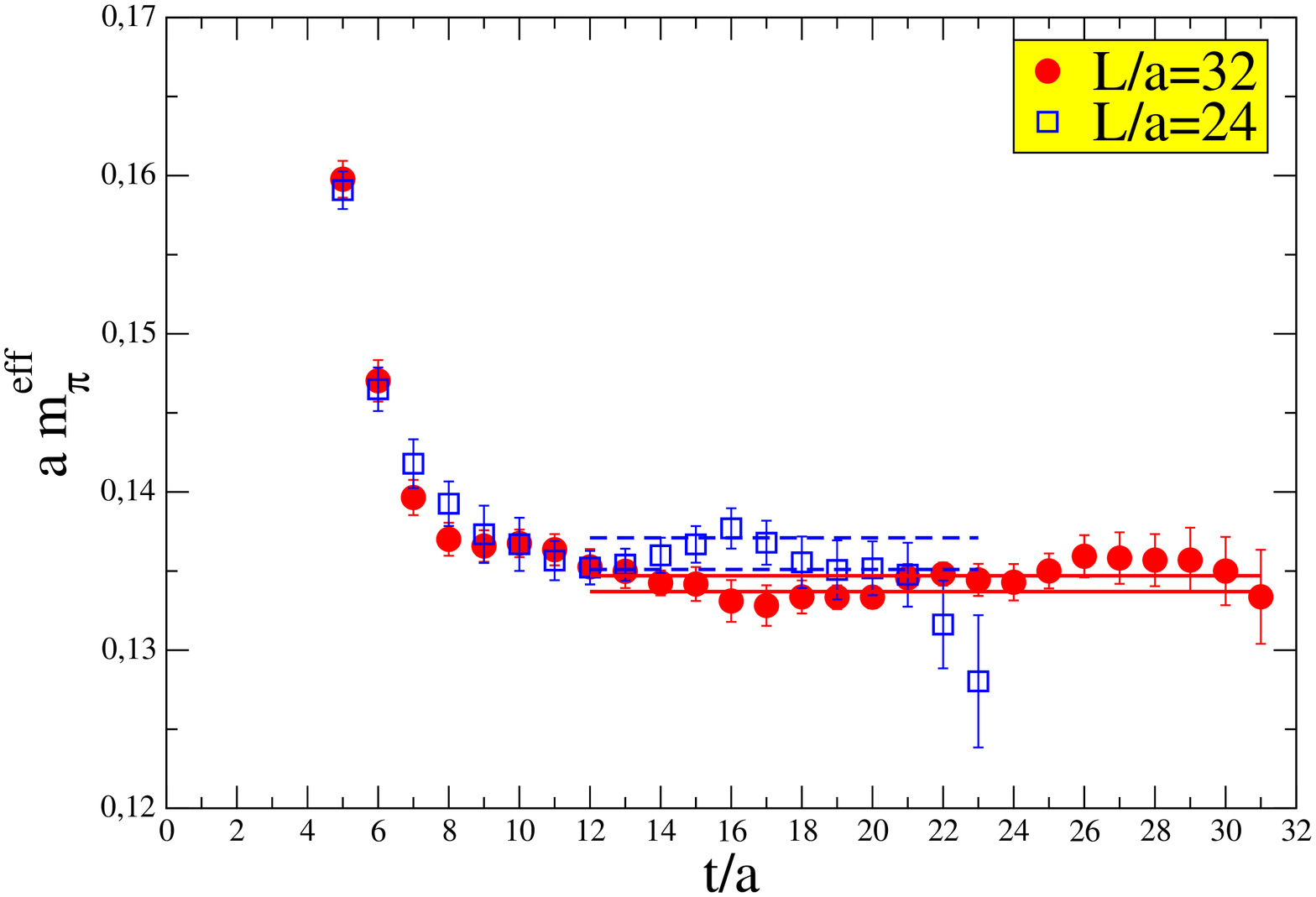}
\hspace{-0.8cm}
\includegraphics[scale=0.3,angle=270,trim=75 0 0 0, clip]{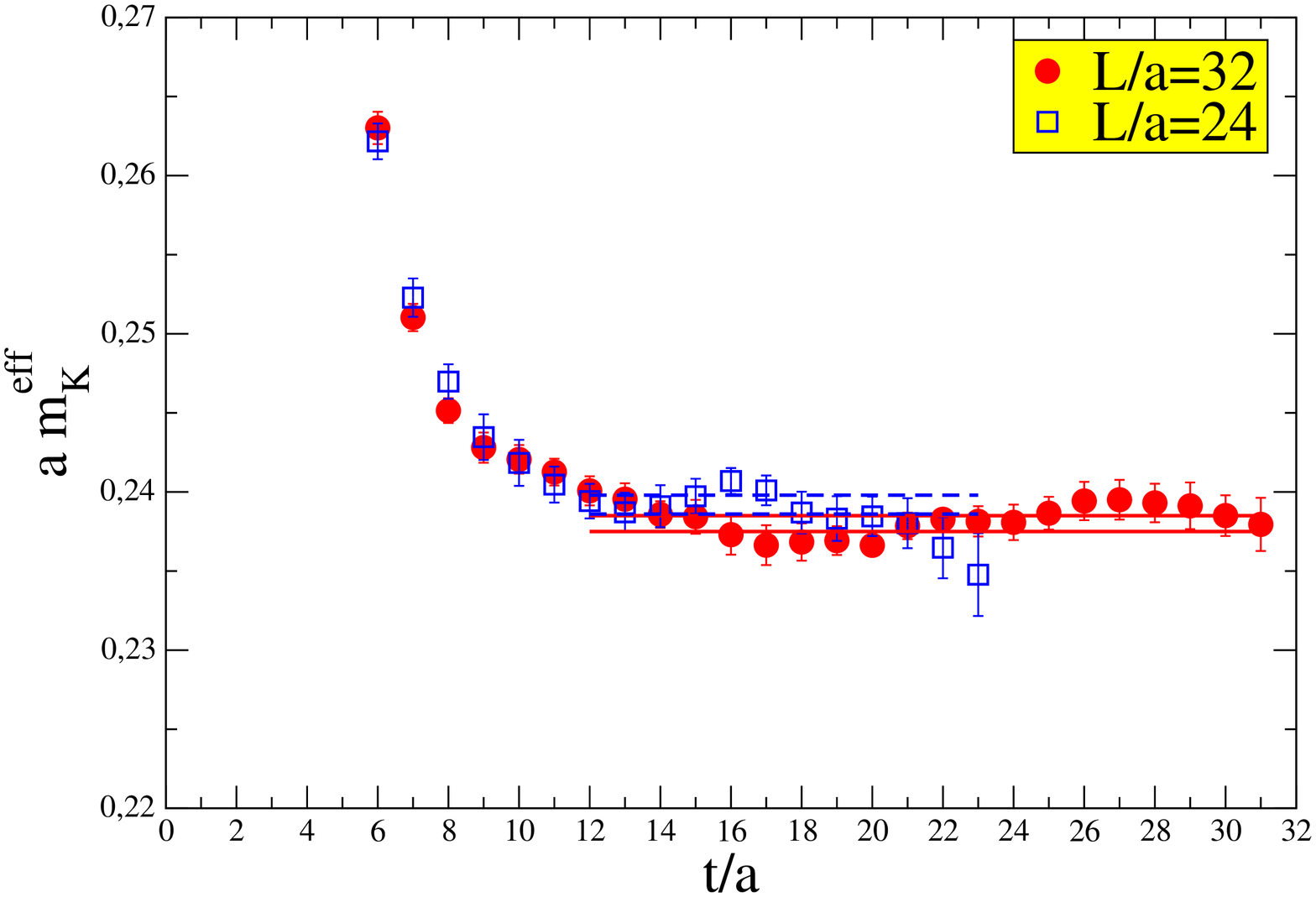}
\vspace{-1.0cm} \\
\includegraphics[scale=0.3,angle=270,trim=75 0 0 0, clip]{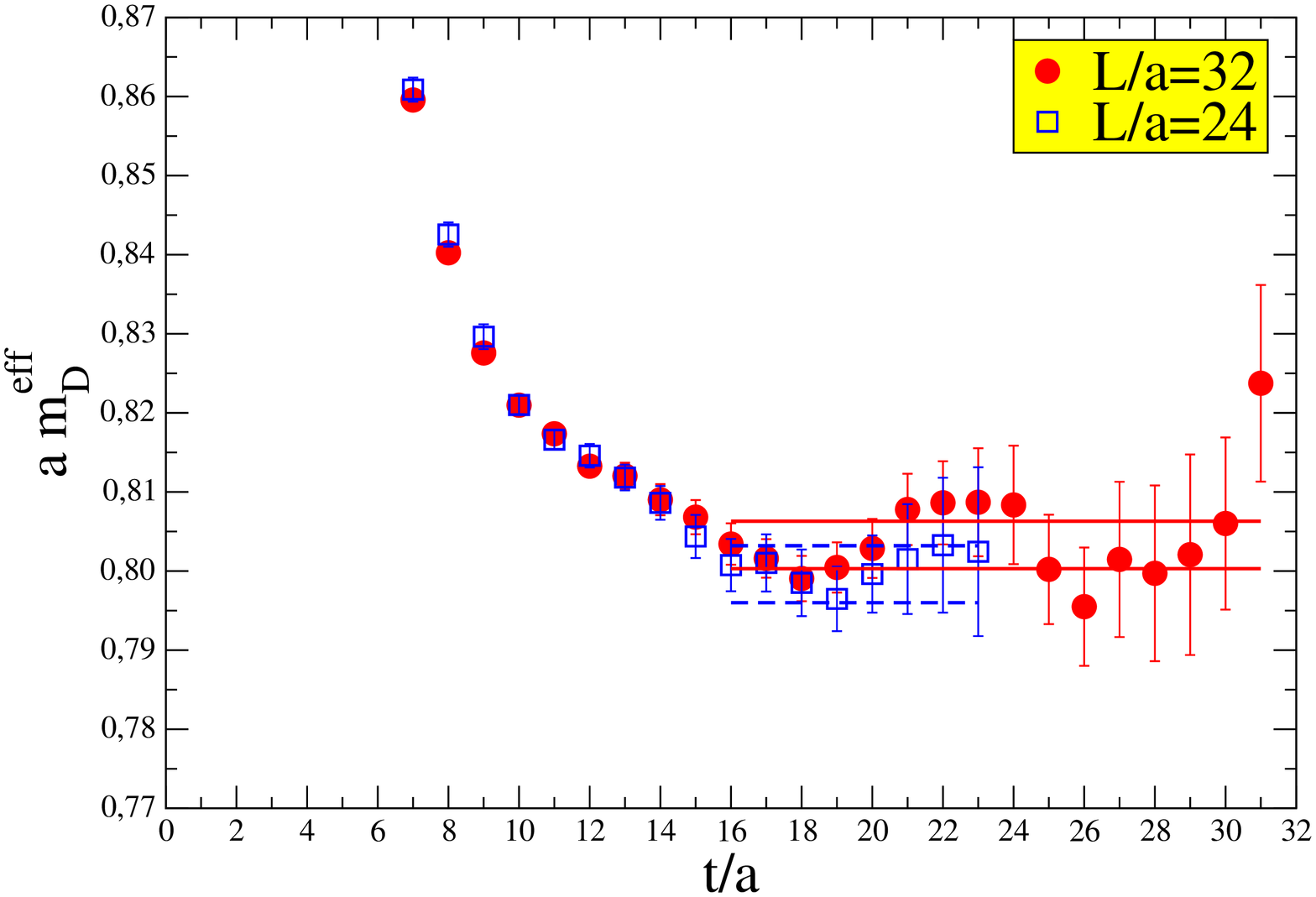}
\hspace{-0.8cm}
\includegraphics[scale=0.3,angle=270,trim=75 0 0 0, clip]{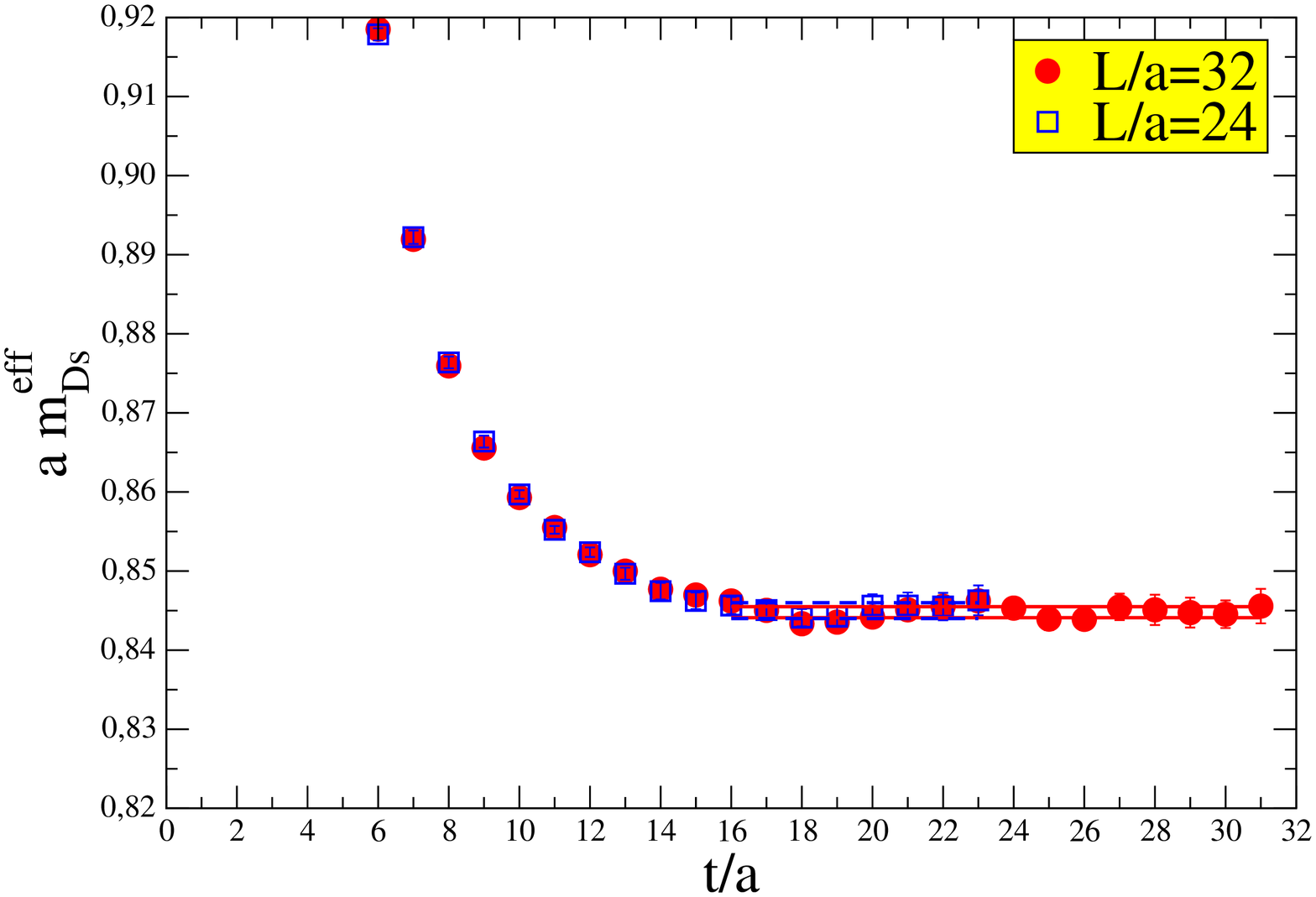}
\vspace{-0.8cm}
\caption{\sl Effective pseudoscalar meson masses $m_{PS}^{eff} (a\,\mu_{sea},
a\,\mu_{val}^{(1)}, a\,\mu_{val}^{(2)})$ as a function of time, in lattice
units, with $\mu_{sea}$ and $\mu_{val}^{(1,2)}$ denoting generically the sea and
valence quark masses respectively. For illustrative purposes the following
choices of quark mass combinations are displayed: $m_{PS}(0.0040,0.0040,0.0040)$
(pion), $m_{PS}(0.0040,0.0040,0.0220)$ (kaon), $m_{PS}(0.0040,0.0040,0.2500)$
($D$-meson), $m_{PS}(0.0040,0.0220,0.2500)$ ($D_s$-meson). In each plot we
compare the effective masses as obtained from the two ensembles $B_1$ and $B_6$,
which correspond to different lattice sizes. Dashed and solid lines represent
the 1-$\sigma$ ranges of the corresponding masses as obtained from the fit of
the two-point correlation functions.}
\label{fig:effmass}
\end{figure}
In order to illustrate the quality of the data, we show in
fig.~\ref{fig:effmass} the effective masses of pseudoscalar mesons, as a
function of the time, for four choices of quark mass combinations, representing
the pion, the kaon, the $D$ and $D_s$ mesons, respectively. The pseudoscalar
masses shown in fig.~\ref{fig:effmass} are extracted from the two point
correlator of the charged pseudoscalar density, together with the matrix element
$\langle 0 | P | PS \rangle$. Both quantities enter eq.~(\ref{eq:fPS}) for
$f_{PS}$. Each plot shows the effective mass obtained from the ensembles $B_1$
and $B_6$, i.e. at $\beta=3.9$, with $a \mu_{sea}=0.0040$ and with two different
lattice sizes, namely $24^3 \times 48$ and $32^3 \times 64$ (see Sec.~3.1 for a
discussion of finite size effects in our analysis).

\section{The pion and kaon decay constants}
\boldmath
\subsection{Combined chiral and continuum extrapolation}
\unboldmath
A good convergence of SU(3)-ChPT is in general not guaranteed in the kaon
sector. As recently pointed out in~\cite{Allton:2008pn} (see
also~\cite{Lellouch:2009fg} for a detailed review on this subject), a safer
approach consists in avoiding the chiral expansion in terms of the strange quark
mass and applying therefore SU(2)-ChPT. The use of SU(2)-ChPT is well motivated
in our analysis, since we simulated $\mu_{s}$ in the range of the physical
strange quark mass, thus having small values of $\mu_l/\mu_s$ (see
Table~\ref{tab:val}). At next-to-leading order (NLO), the SU(2)-ChPT prediction
for the pion decay constant is well known~\cite{Gasser:1983yg},
\beq
\label{eq:fpi}
f_{PS}(\mu_l,\mu_l,\mu_l) = f \cdot \left( 1 - 2\,\xi_{ll}\ln \xi_{ll} +
b\, \xi_{ll} \right) \,,
\eeq
and the corresponding expression for the kaon decay constant
reads~\cite{Allton:2008pn}
\beq
\label{eq:fk}
f_{PS}(\mu_l,\mu_l,\mu_s) = f^{(K)} \cdot \left( 1 - \dfrac{3}{4} \xi_{ll}
\ln \xi_{ll} + b^{(K)}\,\xi_{ll} \right)\, .
\eeq
We are using the notation $(\mu_{sea},\mu^{(1)}_{val},\mu^{(2)}_{val})$ for the
quark mass content of the corresponding meson, and the variables $\xi$'s in
eqs.~(\ref{eq:fpi}) and (\ref{eq:fk}) are expressed in our analysis as a
function of meson masses,\footnote{We use the normalization in which
$f_\pi=130.7\, \mev$.}
\beq
\label{eq:xi}
 \xi_{ij} = \frac{m_{PS}^2(\mu_l,\mu_i,\mu_j)}{(4\pi f)^2}\,.
\eeq
The leading contribution in eq.~(\ref{eq:fpi}) is represented by the low energy
constant (LEC) $f$, which provides the value of the pion decay constant in the
chiral limit, whereas the coefficient $b$ is related to the LEC $\bar l_4$ of
the NLO chiral Lagrangian. Notice that, in the $N_f=2$ theory we are simulating,
i.e. with a quenched strange quark, these constants are independent of the value
of the strange quark mass. In this theory there is actually no distinction
between SU(2) and SU(3) ChPT expansions for pion observables. The LECs $f^{(K)}$
and $b^{(K)}$ entering the SU(2) formula~(\ref{eq:fk}) for the kaon decay
constant, instead, are functions of the (valence) strange quark mass.

In order to perform a combined fit of the data for the pseudoscalar decay
constants at the three values of the lattice spacing, we rely on the Symanzik
expansion of the lattice regularized theory and introduce in the fitting
formulae discretization terms of ${\cal O}(a^2)$ and ${\cal O}(a^2\,\mu_s)$.
Discretization effects of ${\cal O}(a^2\,\mu_l)$, i.e. proportional to the light
quark mass, represent very small contributions that turn out to be invisible in
the fit. Moreover, since the simulated $\mu_s$ masses are all close to the
physical strange quark mass, we can safely linearize the strange mass dependence
of the LECs $f^{(K)}$ and $b^{(K)}$ in eq.~(\ref{eq:fk}) around $m_s^{phys.}$.
We thus write the SU(2)-ChPT fitting formulae for the pion and kaon decay
constants as
\bea
\label{eq:fpifit}
f_{PS}(\mu_l,\mu_l,\mu_l) &=& f \cdot \left( 1 - 2\,\xi_{ll}\ln \xi_{ll} +
b\, \xi_{ll} + A \frac{a^2}{r_0^2}\right)\,, \\
f_{PS}(\mu_l,\mu_l,\mu_s) &=& (f^{(K)}_0 + f^{(K)}_m\, \xi_{ss}) \cdot \nn \\
&& \label{eq:fkfit}
\cdot \left[ 1 - \dfrac{3}{4} \xi_{ll} \ln\xi_{ll} + (b^{(K)}_0 + b^{(K)}_m\,
\xi_{ss})\,\xi_{ll} + (A_a + A_{as}\, \xi_{ss}) \dfrac{a^2}{r_0^2}
\right] \,.
\eea

The fit is performed in units of the Sommer parameter
$r_0$~\cite{Sommer:1993ce}. The values of $r_0/a$ at the three lattice spacings
have been extracted in ref.~\cite{Urbach:2007rt} from the analysis of the static
potential, obtaining
\beq
\label{eq:r0sua}
r_0/a = \left\{4.46(3), \ 5.22(2), \ 6.61(3) \right\}
\eeq
at $\beta = \left\{3.8, \ 3.9, \ 4.05 \right\}$ respectively. The physical value
of $r_0$ in the continuum limit is determined in our analysis by combining the
determination of $r_0\cdot f_\pi$ with the experimental value of the pion decay
constant. This procedure, combined with eq.~(\ref{eq:r0sua}), corresponds to
fixing the lattice scale using $f_\pi$ as physical input.

An important ingredient in the analysis is the study of finite size effects
(FSE). With our simulation setup, the largest FSE are expected in the data of
the ensembles $B_1$ and $C_1$, for which $m_{\pi}\, L \simeq 3.3$ (see
Table~\ref{tab:simdet}). A quantitative estimate of these effects can be
obtained from the comparison of the data of the ensembles $B_1$ and $B_6$, that
only differ in lattice size. This comparison provides consistent results with
the FSE predicted at NLO by SU(2)-ChPT~\cite{Gasser:1987ah,bv}, which are
expressed for the pion and kaon decay constants by
\bea
\label{eq:ffse}
&&f_{PS}(\mu_l,\mu_l,\mu_l;L) = f_{PS}(\mu_l,\mu_l,\mu_l)  \cdot \left[ 1 -
2\,\xi_{ll}\, \tilde g_1(L, \xi_{ll}) \right] \,,\nn\\
&&f_{PS}(\mu_l,\mu_l,\mu_s;L) = f_{PS}(\mu_l,\mu_l,\mu_s)  \cdot \left[ 1 -
\dfrac{3}{4}\,\xi_{ll}\, \tilde g_1(L, \xi_{ll}) \right] \,,
\eea
where the function $\tilde g_1$ is defined for instance in
ref.~\cite{Blossier:2007vv}.\footnote{Note that the finite size corrections in
eq.~(\ref{eq:ffse}) are obtained from the loop contribution of the infinite
volume ChPT predictions in eqs.~(\ref{eq:fpi}) and~(\ref{eq:fk}) by replacing
$\ln \xi$ with the function $\tilde g_1(L, \xi)$.} This correction, which on the
ensembles $B_1$ and $C_1$ corresponds to about 2.5\% for $f_\pi$ and 0.9\% for
$f_K$, has been included in our fit. For a more detailed discussion of FSE in
the pion decay constant see ref.~\cite{Boucaud:2008xu}. In our data for the kaon
decay constant, instead, the differences between the ensembles $B_1$ and $B_6$
are small and at the level of the statistical errors. We obtain $\Delta_{f_K}
\equiv f_K^{L=24}/f_K^{L=32}-1=0.005(4)$, compatible with zero. Also the NLO
partially quenched ChPT prediction for $\Delta_{f_K}$~\cite{bv} is similarly
small ($-0.006$), at the level of our statistical error.

\begin{figure}[t]
\begin{center}
\vspace{-0.85cm}
\includegraphics[scale=0.4,angle=270]{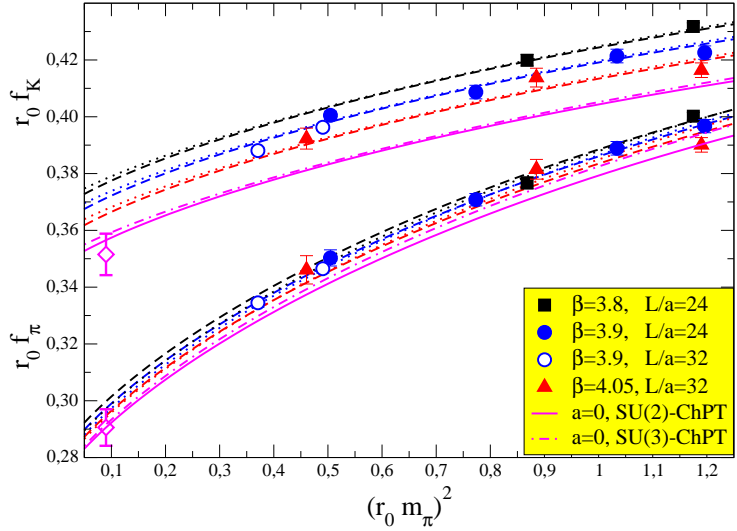} \\
\end{center}
\vspace{-1.0cm}
\caption{\sl Lattice results for $r_0 f_{\pi}\equiv r_0
f_{PS}(\mu_l,\mu_l,\mu_l)$ and $r_0 f_{K}\equiv r_0 f_{PS}(\mu_l,\mu_l,\mu_s)$
as a function of the pion mass square $(r_0 m_\pi)^2 \equiv (r_0
m_{PS}(\mu_l,\mu_l,\mu_l))^2$. For the kaon, we display data with $\mu_s$ fixed
to the simulated mass that corresponds to a reference meson mass $r_0
m_{PS}(\mu_l,\mu_s,\mu_s)\simeq 1.63$.  The SU(2)- (SU(3)-) ChPT extrapolation
to the physical pion mass is represented at fixed lattice spacing by the dashed
(dotted) curves, and in the continuum limit by the solid (dashed-dotted) curve.
Our results for the physical values of the pion and kaon decay constants,
obtained from SU(2)-ChPT, are illustrated by diamonds in the plot. In the kaon
case an interpolation to the physical strange quark mass is performed.}
\vspace{-0.3cm}\label{fig:fKSU3vsSU2}
\end{figure}

The values of the fit parameters and the results for $r_0$, $f_K$ and
$f_K/f_\pi$ are collected in Tables~\ref{tab:parSU2} and~\ref{tab:resSU3SU2},
respectively.
\begin{table}[t]
\begin{center}
\begin{tabular}{|c|c|c|c|c|c|c|c|c|}
\hline
$r_0\cdot f$ & $r_0\cdot f^{(K)}_0$ & $r_0 \cdot
f^{(K)}_m$ & $b$ & $b^{(K)}_0$ & $b^{(K)}_m$ & $A$ & $A_a$ & $A_{as}$
\\ \hline
$0.271(6)$ & $0.305(6)$ & $0.18(1)$ & $-0.25(13)$ &
$0.5(1)$   & $-1.9(4)$  & $0.7(5)$  & $0.7(4)$    & $2.1(6)$ \\
$0.274(9)$ & $0.312(9)$ & $0.18(2)$ & $-0.33(15)$ &
$0.4(2)$   & $-1.7(5)$  & $0.5(7)$  &  $0.3(7)$   & $3(1) $\\
\hline
\end{tabular}
\end{center}
\vspace*{-0.5cm}
\caption{\sl Values of the SU(2) fit parameters of eqs.~(\ref{eq:fpifit})
and~(\ref{eq:fkfit}), as obtained by including (first row) or excluding (second
row) the data at $\beta=3.8$. Quoted errors are statistical plus fitting
errors.}
\label{tab:parSU2}
\end{table}
\begin{table}[t]
\begin{center}
\begin{tabular}{|c|c||c|c|c|c||}\cline{3-6}
\multicolumn{2}{c||}{} & 
$r_0 \,[\gev^{-1}]$ & $f_K\, [\mev]$ & $f_K/f_\pi$ & $\chi^2/dof$ \\ \hline
SU(2)-ChPT & incl. $\beta=3.8$& $2.22(5)$ & $158.1(8)$ & $1.210(6)$ & 
$\{11/8;\, 40/30\}$\\
           & excl. $\beta=3.8$& $2.25(7)$ & $158.9(1.4)$ & $1.216(11)$
&$\{7/6;\, 35/22\}$\\
\hline
SU(3)-ChPT & incl. $\beta=3.8$& $2.23(5)$ & $158.2(6)$ & $1.210(5)$ & $61/42$\\
           & excl. $\beta=3.8$& $2.28(7)$ & $158.0(0.8)$ & $1.209(6)$ & $54/32$\\
\hline
\end{tabular}
\end{center}
\vspace{-0.4cm}
\caption{\sl Values of $r_0$, $f_K$ and $f_K/f_\pi$ as obtained from SU(2)- and
SU(3)-ChPT by including or excluding the data at $\beta=3.8$. For each fit, the
chi-squared per degree of freedom, $\chi^2/dof$, is also given in the last
column. For fits based on SU(2)-ChPT, the two values of $\chi^2/dof$ correspond
to the fit of $f_\pi$ and $f_K$ respectively. Quoted errors are statistical plus
fitting errors.}
\label{tab:resSU3SU2}
\end{table}
The extrapolation to the physical up/down quark mass and the interpolation to
the physical strange mass has been performed by inserting in
eqs.~(\ref{eq:fpifit}) and~(\ref{eq:fkfit}) $\xi_{ll}=(m_\pi^{phys.})^2/(4 \pi
f)^2$ and $\xi_{ss}=(2\,(m_K^{phys.})^2-(m_\pi^{phys.})^2)/(4 \pi f)^2$, where
$m_\pi^{phys.}$ and $m_K^{phys.}$ are the experimental pion and kaon masses. In
both Tables, we show the results of our fits when we take the data at
$\beta=3.8$ into account and when we leave them out. As can be seen, the values
of the fit parameters, as well as those of the decay constants, are found to be
well consistent in the two cases. In the following, therefore, we will consider
for $f_K$ and $f_K/f_\pi$ only the predictions obtained by including the
$\beta=3.8$ data. From the results given in Tables~\ref{tab:parSU2}
and~\ref{tab:resSU3SU2}, one can also derive our prediction for the pion decay
constant in the chiral limit, $f$, and the LEC $\bar l_4 = b/2 + 2\ln(4 \pi
f/m_{\pi^+})$. We obtain the values $f=121.7(1) \mev$ and $\bar l_4=4.66(6)$,
which are in good agreement with the results of the scaling analysis performed
by our Collaboration in~\cite{Dimopoulos:2008sy}, $f=121.66(7)(26) \mev$ and
$\bar l_4=4.59(4)(13)$. The quality of the fit for the combined chiral and
continuum extrapolation of the pion and kaon decay constant is illustrated in
fig.~\ref{fig:fKSU3vsSU2}.

As an alternative to the SU(2)-ChPT approach, we have also considered the
expansion valid for a small strange quark mass, fitting both the pion and the
kaon decay constants using SU(3)-ChPT. The relevant expression, valid for the
partially quenched $N_f=2$ theory at NLO, is~\cite{Sharpe:1997by}
\beq
\label{eq:fSU3}
f_{PS}(\mu_l,\mu_l,\mu_{s}) = f  \cdot \left[ 1 - \dfrac{3}{4} \xi_{ll}
\ln\xi_{ll} - \dfrac{\xi_{ll}}{4}\ln \xi_{ss} - \xi_{ls}\ln\xi_{ls}+ b_{ll}
\xi_{ll} + b_{ss} \xi_{ss} + (B_a + B_{as}\, \xi_{ss}) \dfrac{a^2}{r_0^2}
\right] \,,
\eeq
where, as in the SU(2) case, we have also included in the fit discretization
terms of ${\cal O}(a^2)$ and ${\cal O}(a^2\,\mu_s)$ as well as finite size
corrections~\cite{bv}.

The results obtained from the SU(3)-ChPT analysis are given in
Table~\ref{tab:resSU3SU2} and shown in fig.~\ref{fig:fKSU3vsSU2}, and are found
in very good agreement with those obtained from the SU(2) fit. A more careful
analysis suggests, however, that the SU(3)-ChPT fit is less robust than the one
based on SU(2). In SU(2)-ChPT, at NLO, one obtains a good fit of the data by
expressing the chiral formulae~(\ref{eq:fpifit}) and  (\ref{eq:fkfit}) either in
terms of meson masses, as performed here with $\xi_{ij}$ defined as in
eq.~(\ref{eq:xi}), or in terms of quark masses, i.e. with
$\xi_{ij}=B\,(\mu_i+\mu_j)/(4\pi f)^2$, where $B$ is the LEC entering at LO in
the chiral Lagrangian. In SU(3)-ChPT, instead, the NLO formula expressed in
terms of meson masses provides a good description of the lattice  data, but fits
performed in terms of quark masses require in the kaon sector the inclusion of
NNLO terms, as we already found in~\cite{Blossier:2007vv}. This means that the
replacement of quark with meson masses effectively resums higher order chiral
contributions, actually improving the fit based on NLO SU(3)-ChPT of the
pseudoscalar decay constant beyond $m_{PS} \sim 450 \,\mev$. A similar result
was found in ref.~\cite{Noaki:2008iy}.

\boldmath
\subsection{Results for $f_K$ and $f_K/f_\pi$}
\unboldmath
As seen in the previous section, lattice data in the kaon sector could be
analysed by means of either SU(2)- or SU(3)-ChPT, leading to almost identical
results, see Table~\ref{tab:resSU3SU2}. The results we quote for $f_K$ and
$f_K/f_{\pi}$ are those obtained from SU(2). The errors quoted in
Table~\ref{tab:resSU3SU2} are statistical plus fitting errors from the combined
chiral and continuum extrapolation. We now discuss how we evaluate other
systematic errors.

Since we have simulated at three values of the lattice spacing and on our
coarsest lattice ($\beta=3.8$) we have data for only two values of the light
quark mass, we include in our final results a systematic uncertainty due to
residual discretization effects. The leading discretization errors in our
determination of the light meson decay constants are expected to be of ${\cal
O}(a^2\, \Lambda_{QCD}^2)$. This na\"ive expectation is roughly confirmed by the
results of our fit. On our finest lattice, for instance, with $\beta=4.05$ and
$a\simeq 0.07\,\fm$, one has $a^2\, \Lambda_{QCD}^2 \simeq 1\div 2\%$ and we
find that the difference between the values taken by the kaon decay constant on
this lattice and its estimate in the continuum limit is approximately $2.6\%$
(this difference turns out to be slightly larger, about $2.8 \%$, at the
reference mass $r_0 m_{PS} (\mu_l, \mu_s, \mu_s)=1.63$ for which results are
displayed in fig.~\ref{fig:fKSU3vsSU2}). We conservatively assume an error of
50\% in the continuum extrapolation starting from the data on our finest
lattice, thus adding to our final results for $f_K$ and $f_K/f_\pi$ a relative
systematic error of 1.3\% (half of the difference between the values at
$\beta=4.05$ and the continuum estimates).

In the present analysis, FSE corrections have been implemented by using the
predictions of NLO ChPT, as discussed in the previous section. Besides the
direct comparison of this theoretical estimate with the data available on the
two lattices $B_1$ and $B_6$, where $m_\pi\,L$ varies from $3.3$ to $4.3$, an
additional indication that these corrections are under control is provided by
the compatibility between the results for $f_\pi$ determined here, by treating
the FSE with NLO ChPT, and those obtained in~\cite{Boucaud:2008xu} by using the
resummed formulae of ref.~\cite{Colangelo:2005gd}. For the kaon decay constant,
FSE are found at the level of the statistical errors at most. A fit performed
without including this correction provides a result for the kaon decay constant
which is lower by about 0.7\%. We conservatively include this difference in the
systematic error of $f_K$ as an estimate of the uncertainty due to FSE.

\begin{figure}[t]
\begin{center}
\includegraphics[scale=0.4,angle=270]{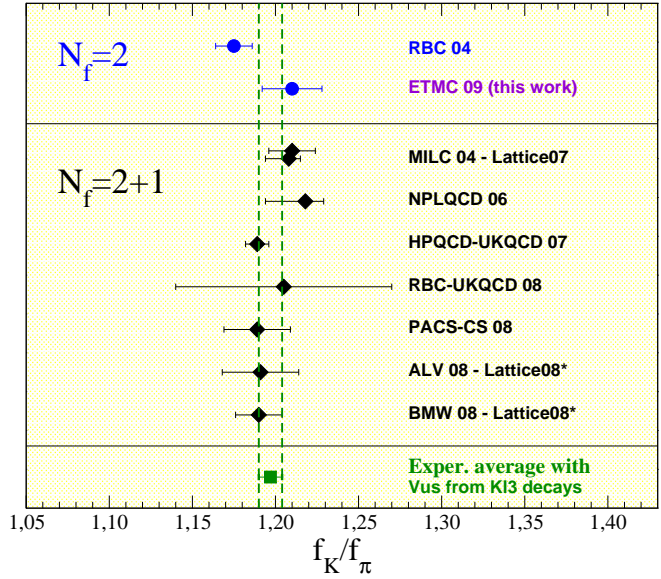}
\end{center}
\vspace{-1.0cm}
\caption{\small\sl  Lattice QCD determinations of the ratio $f_K/f_\pi$ obtained
from simulations with $N_f=2$~\cite{Aoki:2004ht} and
$N_f=2+1$~\cite{Allton:2008pn,Lellouch:2009fg},~\cite{Aoki:2004ht}-\cite{
Aubin:2008ie} dynamical quarks. A star in the legend denotes preliminary
results. The results are also compared with the experimental average of
$f_K/f_\pi$ obtained by using for $V_{us}$ the determination from $K_{\ell 3}$
decays~\cite{Antonelli:2008jg}.}
\label{fig:fkfpiunq}
\end{figure}

The only uncertainty which cannot be reliably estimated within our $N_f=2$
simulation, is the error due to the quenching of the strange quark. The good
agreement observed among the recent $N_f=2$ and $N_f=2+1$ lattice determinations
of $f_K$ and $f_K/f_\pi$ (see fig.~\ref{fig:fkfpiunq}) suggests, however, that
such an effect is smaller than the other systematic uncertainties estimated
above. The ETM Collaboration is planning to investigate directly the effect of
the quenching of both the strange and charm quarks through simulations with
$N_f=2+1+1$, which are currently in progress~\cite{Baron:2008xa}. For a more
extensive discussion of the various sources of lattice systematic uncertainties
see ref.~\cite{Jansen:2008vs}.

Our final results for the kaon decay constant and the ratio $f_K/f_\pi$ are then
\beq
f_K =  158.1(0.8)(2.0)(1.1)\ \mev\, \qquad , \qquad 
f_K/f_\pi = 1.210(6)(15)(9)\,,
\eeq
where the first error comes from statistics and chiral extrapolation, the second
from the estimate of residual discretization effects and the third from the
uncertainty on FSE corrections. By combining the errors in quadrature, we
finally obtain
\beq
f_K =  158.1(2.4)\ \mev\, \qquad , \qquad f_K/f_\pi = 1.210(18)\,.
\label{eq:fKfinal}
\eeq

Our result for the ratio $f_K/f_\pi$ is compared in fig.~\ref{fig:fkfpiunq} with
other unquenched lattice determinations, performed with either $N_f=2$ or
$N_f=2+1$ dynamical quarks, as well as with the experimental average of
$f_K/f_\pi$ obtained by using for $V_{us}$ the determination from $K_{\ell 3}$
decays~\cite{Antonelli:2008jg}. Our result turns out to be in very good
agreement with the latter determination, as well as with most of the $N_f=2$ and
$N_f=2+1$ lattice results.

Alternatively, our result for $f_K/f_\pi$ can be combined with the experimental
measurement of $\Gamma(K \to \mu \bar \nu_\mu (\gamma))/\Gamma(\pi \to \mu \bar
\nu_\mu (\gamma))$~\cite{Antonelli:2008jg} to get a determination of the ratio
$\vert V_{us}\vert/\vert V_{ud}\vert$~\cite{Marciano:2004uf}. We obtain
\beq
\label{eq:vuds}
\vert V_{us}\vert/\vert V_{ud}\vert= 0.2281(5)(35)\, ,
\eeq
where the first error is the experimental one and the second is the theoretical
error coming from the uncertainty on $f_K/f_\pi$. Eq.~(\ref{eq:vuds}), combined
with the determination $\vert V_{ud} \vert= 0.97425(22)$~\cite{Hardy:2008gy}
from nuclear beta decays, yields the estimate
\beq
\label{eq:vus-kl3}
\vert V_{us}\vert= 0.2222(5)(34) \, ,
\eeq
in good agreement, though with a larger error, with the value extracted from
$K_{\ell 3}$ decays, $\vert V_{us} \vert= 0.2246(12)$~\cite{Antonelli:2008jg}.
Eq.~(\ref{eq:vus-kl3}) and the value of $\vert V_{ud} \vert$ quoted above leads
to
\beq
\vert V_{ud}\vert^2 + \vert V_{us}\vert^2 + \vert V_{ub}\vert^2 -1 =
-1.5 (1.6)\cdot 10^{-3} \, .
\eeq
in good agreement with the unitarity constraint of the CKM matrix.

\boldmath
\section{The $D$ and $D_s$ decay constants}
\subsection{Combined chiral and continuum extrapolation}
\unboldmath

In order to determine the $D$ and $D_s$ meson decay constants we essentially
proceed as in the kaon sector. We analyse simultaneously data at the three
values of the lattice spacing and perform for the pseudoscalar decay constants
combined fits of the meson mass dependence and of discretization effects. The
simulated values $\mu_c$ of the charm quark mass are close to the physical charm
quark mass ($0.8\, m_c^{phys.} \simle \mu_c \simle 1.5\, m_c^{phys.}$), so that
the interpolation to the physical value is short and smooth. From the comparison
of the data of the ensembles $B_1$ and $B_6$ we also find that FSE are
negligible for the $D_s$ decay constants, and they are at the level of the
statistical error or smaller for $f_D$. On the other hand, discretization errors
induced by the charm quark mass have to be taken carefully into account in the
fit, being parametrically of $ {\cal O}(a^2\, \mu_c^2)$, i.e. approximately $5
\div 10\%$ in our simulation.

The functional forms describing the mass dependence of the decay constants
assumed to fit the data in the $D$ and $D_s$ sectors are those predicted by
HMChPT~\cite{Sharpe:1995qp}. We consider the SU(2) version of the theory, as in
the case of the kaon sector, where the strange quark is not required to satisfy
chiral symmetry, but it is considered heavy enough to justify an expansion in
powers of $\mu_l/\mu_s$. For comparison, we have also investigated the
predictions of SU(3)-HMChPT where, instead, the strange quark is required to
satisfy the same chiral symmetry of the light up and down quarks. As we will see
below, in the $D$-meson sector the SU(2)-HMChPT approach turns out to work
significantly better than the one based on SU(3).

Within the SU(2)-HMChPT analysis, we extract $f_D$ and $f_{D_s}$ by considering
two different procedures. In the first one we fit the two following combinations
of meson masses and decay constants:
\beq
f_{D_s} \sqrt{m_{D_s}}\,, \qquad  R \equiv \dfrac{f_{D_s} \sqrt{m_{D_s}}}{f_D
\sqrt{m_D}}\,, \qquad  \rm{(Fit ~ I)} \ .
\label{eq:directs}
\eeq
We find, in particular, that the advantage of introducing the ratio $R$ is that
discretization effects largely cancel in the ratio. In the second approach we
consider the previous quantities divided by the light decay constants, i.e. we
fit the ratios
\beq
R_1 \equiv \dfrac{f_{D_s} \sqrt{m_{D_s}}}{f_K}\,, \qquad  R_2 \equiv
\dfrac{f_{D_s} \sqrt{m_{D_s}}}{f_K} \times \dfrac{f_\pi}{f_D \sqrt{m_D}}\,.
\qquad \rm{(Fit ~ II)}
\label{eq:ratios}
\eeq
Here, the advantage of the ratio $R_2$ is that it exhibits a quite smooth chiral
behaviour. The comparison of the results obtained for $f_D$, $f_{D_s}$ and
$f_{D_s}/f_D$ in the two cases, Fit I and Fit II, will provide an estimate of
the systematic uncertainty due to the chiral extrapolation.

In all quantities entering eqs.~(\ref{eq:directs}) and~(\ref{eq:ratios}), the
$D$-mesons decay constants are multiplied by the square roots of the
corresponding meson masses, in order to reconstruct the observables that remain
finite in the infinite mass limit. The Heavy Quark Effective Theory predicts in
fact for a Heavy($H$)-light($l$) meson an expansion of the form $f_{Hl}
\sqrt{m_{Hl}} = A + B/m_{Hl} +{\cal O}(1/m_{Hl}^2)$, up to small radiative
corrections. Though the heavy quark expansion is known to be slowly convergent
in the charm mass region, in our analysis we can safely assume such a behaviour
for the $D$ mesons since only a short interpolation of the lattice data to the
physical charm quark mass is needed. Moreover, since the contribution of the
sub-leading $1/m_{Hl}$ correction in this interpolation is small, we can safely
account for the dependence on the light meson masses only in the leading term,
by using the prediction of HMChPT.

We obtain the SU(2)-HMChPT functional forms for the quantities in
eqs.~(\ref{eq:directs}) and~(\ref{eq:ratios}) by expanding the corresponding
SU(3)-HMChPT predictions in powers of $\mu_l/\mu_s$, and reabsorbing the strange
quark mass dependence in the SU(2) LECs.\footnote{The same procedure allows to
obtain the SU(2)-ChPT expression~(\ref{eq:fk}) of the kaon decay constant $f_K$
from the SU(3) prediction of eq.~(\ref{eq:fSU3}).} The SU(3)-HMChPT prediction
for $f_{D_s} \sqrt{m_{D_s}}$, valid in the partially quenched $N_f=2$
theory~\cite{Sharpe:1995qp}, is given by
\beq
f_{D_s} \sqrt{m_{D_s}} = \dfrac{C_1}{r_0^{3/2}} \left[ 1 - \dfrac{1+3 
g_c^2}{2} \left( 2 \xi_{ls} \ln \xi_{ls} + \dfrac{\xi_{ll}-2\xi_{ls}}{2} \ln
\xi_{ss}\right)+ C_2 \xi_{ll} + C_3 \xi_{ss}\right] +\dfrac{C_4}{r_0^{5/2} m_{D_s}}\,,
\label{eq:fDsSU3}
\eeq
where the parameter $g_c$ is related to the $g_{D^*D\pi}$ coupling by
$g_{D^*D\pi}= (2 \sqrt{m_D m_{D^*}}/f_\pi)\, g_c$. Since eq.~(\ref{eq:fDsSU3})
does not contain logarithms of the pion mass (i.e. $\ln \xi_{ll}$), one finds
that its expansion in powers of $\mu_l/\mu_s \simeq \xi_{ll}/\xi_{ss}$ leads to
an SU(2) chiral expression for the $D_s$ decay constant which is free of chiral
logarithms at NLO. As in the light meson case, we also include in the fitting
formula discretization terms of ${\cal O}(a^2)$, in order to take simultaneously
into account the lattice artefacts. We observe, in this respect, that the
Symanzik expansion of $f_D$ and $f_{D_s}$ contains at ${\cal O}(a^2)$
discretization terms depending on the charm quark mass either linearly or
quadratically, with the leading contribution expected from terms of ${\cal
O}(a^2\,\mu_c^2)$. The limited set of data available in our analysis, however,
does not allow us to fit both these dependencies separately. We parameterize
these discretization effects in terms of the meson masses, and we thus introduce
in the fitting formula only a term proportional to $a^2\, m^2_{D_s}$. We have
also tried an alternative fit where the charm mass dependent discretization term
is taken to be proportional to $a^2\, m_{D_s}$ instead of $a^2\, m^2_{D_s}$, and
we obtain completely consistent results. Thus, we use as our fitting formula for
$f_{D_s} \sqrt{m_{D_s}}$ the expression
\beq
f_{D_s} \sqrt{m_{D_s}} = \dfrac{D_1}{r_0^{3/2}} \left[ 1 +D_2 \xi_{ll} + \left(
D_a + D_{as} \xi_{ss} \right) \dfrac{a^2}{r_0^2} + D_{ah}\, a^2 m^2_{Ds}
 \right] +\dfrac{D_3}{r_0^{5/2} m_{D_s}}\,,
\label{eq:fDsSU2}
\eeq
where the coefficients $D_1$ and $D_2$, which depend on the strange quark mass,
are expressed in the fit as linear functions of this mass:
\beq
 D_i=D_{i,0} + D_{i,m}\, \xi_{ss}\,,
\label{eq:lin}
\eeq
with $i=1,2$.

The SU(2)-HMChPT prediction for the ratio $R$, defined in
eq.~(\ref{eq:directs}), is straightforwardly obtained by dividing
eq.~(\ref{eq:fDsSU2}) by the HMChPT expression for $f_{D} \sqrt{m_{D}}$, which
is provided by the SU(3)-HMChPT formula of eq.~(\ref{eq:fDsSU3}) with
$\mu_s=\mu_l$. Thus, we assume for the ratio $R$ the expression
\beq
R = D^{\prime}_1 \left[ 1 +  \dfrac{3}{4} \left(1+3 g_c^2\right) \xi_{ll}
\ln \xi_{ll} + D^{\prime}_2 \xi_{ll} + \left(D^{\prime}_a + D^{\prime}_{as}
\xi_{ss} \right) \dfrac{a^2}{r_0^2} + D^{\prime}_{ah}\, a^2 m^2_{Ds} \right] +\dfrac{D^{\prime}_3}{r_0 m_{D_s}}\,,
\label{eq:RSU2}
\eeq
where the coefficients $D^{\prime}_1$ and $D^{\prime}_2$ are expanded as linear
functions of the strange quark mass as in eq.~(\ref{eq:lin}).

We find that the HMChPT parameter $g_c$ cannot be determined from the fit, which
is almost insensitive to it. It is thus constrained to the experimental value
$g_c = 0.61(7)$~\cite{PDG,Anastassov:2001cw} which is in good agreement with a
recent unquenched lattice determination,  $g_c =
0.71(7)$~\cite{Becirevic:2009xp}.

\begin{table}[t]
\begin{center}
\begin{tabular}{|c|c|c|c|c|c|c|c|}
\hline 
$D_{1,0}$ & $D_{1,m}$ & $D_{2,0}$ & $D_{2,m}$ & $D_3$ & $D_a$ & $D_{as}$  &
$D_{ah}$ \\ \hline
$1.62(9)$ & $0.78(7)$ & $0.4(2)$ & $-0.7(5)$ & $-2.9(3)$ & $-0.2(4)$ & $-2(1)$ &
$0.13(3)$ \\
$1.54(9)$ & $0.9(1)$ & $0.7(2)$ & $-1.3(5)$ & $-2.6(3)$ & $-0.6(7)$ & $-3(2)$ &
$0.16(2)$ \\\hline\hline
$D^{\prime}_{1,0}$ & $D^{\prime}_{1,m}$ & $D^{\prime}_{2,0}$ &
$D^{\prime}_{2,m}$ & $D^{\prime}_3$ & $D^{\prime}_a$ & $D^{\prime}_{as}$  &
$D^{\prime}_{ah}$  \\ \hline
$1.0(1)$ & $1.1(2)$ & $1.9(4)$ & $-2.1(9)$ & $0.6(4)$ & $-1.4(9)$ & $0(1)$ &
$0.07(5)$ \\
$1.0(1)$ & $1.2(2)$ & $2.1(6)$ & $-3(1)$ & $0.5(6)$ & $-1(1)$ & $1(2)$ &
$0.07(7)$ \\\hline
\end{tabular}

\vspace*{0.5cm}

\begin{tabular}{|c|c|c|c|c|c|c|c|}
\hline 
$P_{1,0}$ & $P_{1,m}$ & $P_{2,0}$ & $P_{2,m}$ & $P_3$ & $P_a$ & $P_{as}$ &
$P_{ah}$ \\ \hline
$4.9(2)$ & $0.7(2)$ & $0.7(2)$ & $0.1(5)$ & $-7.2(8)$ & $-0.1(4)$ & $-3(1)$ &
$0.11(2)$ \\
$4.7(2)$ & $0.8(3)$ & $0.9(2)$ & $-0.5(5)$ & $-6.8(7)$ & $-0.2(6)$ & $-4(2)$ &
$0.12(2)$ \\\hline\hline
$P^{\prime}_{1,0}$ & $P^{\prime}_{1,m}$ & $P^{\prime}_{2,0}$ &
$P^{\prime}_{2,m}$ & $P^{\prime}_3$ & $P^{\prime}_a$ & $P^{\prime}_{as}$ &
$P^{\prime}_{ah}$  \\ \hline
$0.9(1)$ & $0.4(1)$ & $0.9(5)$ & $-2(1)$ & $0.4(4)$ & $-2(1)$ & $-1(2)$ &
$0.08(5)$ \\
$0.9(1)$ & $0.4(1)$ & $0.9(6)$ & $-3(2)$ & $0.3(5)$ & $-1(2)$ & $1(3)$ &
$0.06(7)$ \\\hline
\end{tabular}
\end{center}
\vspace{-0.5cm}
\caption{\sl Values of the SU(2)-HMChPT fit parameters from Fit I of
eqs.~(\ref{eq:fDsSU2}) and~(\ref{eq:RSU2}) (upper table) and from Fit II of
eq.~(\ref{eq:R1R2}) (lower table), as obtained by including (first row) or
excluding (second row) the data at $\beta=3.8$. Quoted errors are statistical
plus fitting errors.}
\label{tab:parSU2HM}
\end{table}
\begin{table}[t]
\begin{center}
\begin{tabular}{|c|c||c|c|c|c||}\cline{3-6}
\multicolumn{2}{c||}{} & $f_D\, [\mev]$ & $f_{D_s}\, [\mev]$ &
$f_{D_s}/f_D$ & $\chi^2/dof$ \\ \hline
SU(2)-HMChPT & incl. $\beta=3.8$ & 
$195(6)$ & $242(3)$ & $1.24(3)$ & $\{93/136;61/136\}$ \\
   Fit I     & excl. $\beta=3.8$ &
$194(8)$ & $239(5)$ & $1.23(4)$ & $\{25/96;49/96\}$ \\ \hline
SU(2)-HMChPT & incl. $\beta=3.8$ &
$199(6)$ & $246(3)$ & $1.24(3)$ & $\{146/136;69/136\}$ \\
   Fit II    & excl. $\beta=3.8$ &
$195(8)$ & $243(5)$ & $1.24(5)$ & $\{24/96;39/96\}$ \\ \hline
SU(3)-HMChPT & incl. $\beta=3.8$ & 
$197(6)$ & $239(3)$ & $1.22(2)$ & $371/179$ \\ \hline
\end{tabular}
\end{center}
\vspace{-0.4cm}
\caption{\sl Values of $f_D$, $f_{D_s}$ and $f_{D_s}/f_D$ as obtained from the
SU(2)-HMChPT Fits I and II by including or excluding data at $\beta=3.8$. We
also show in the last row the results obtained by fitting both $f_{D_s}
\sqrt{m_{D_s}}$ and $f_{D} \sqrt{m_{D}}$ with their common SU(3)-HMChPT
functional form. For each fit, the chi-squared per degree of freedom is given in
the last column. For fits based on SU(2)-HMChPT, two values of $\chi^2/dof$ are
displayed, corresponding to $f_{D_s}$ and $R$ for Fit I or $R_1$ and $R_2$ for
Fit II. Quoted errors are statistical plus fitting errors.}
\label{tab:resSU2HM}
\end{table}
The values of the coefficients $D_i$ and $D^{\prime}_i$ as resulting from the
fits of eqs.~(\ref{eq:fDsSU2}) and~(\ref{eq:RSU2}) are collected in
Table~\ref{tab:parSU2HM}. Using these results and inserting in
eqs.~(\ref{eq:fDsSU2}) and~(\ref{eq:RSU2}) the experimental values of the
relevant meson masses we obtain for $f_D$, $f_{D_s}$ and $f_{D_s}/f_D$ the
results given in Table~\ref{tab:resSU2HM} labelled as SU(2)-HMChPT, Fit I. As in
the light meson case, we show in the Tables the results obtained by including or
excluding the data at $\beta=3.8$. The values of the fit parameters are
consistent in the two cases and the results for the decay constants are
essentially equal.

The alternative approach we considered to determine the $D$ and $D_s$ meson
decay constants is based on the study of the ratios $R_1$ and $R_2$ defined in
eq.~(\ref{eq:ratios}). The SU(2)-HMChPT predictions for these ratios are easily
obtained by dividing the expressions~(\ref{eq:fDsSU2}) and~(\ref{eq:RSU2}) for
$f_{D_s} \sqrt{m_{D_s}}$ and $R$ by the SU(2)-ChPT predictions~(\ref{eq:fpi})
and~(\ref{eq:fk}) for $f_{\pi}$ and $f_K$, corrected for FSE as in
eq.~(\ref{eq:ffse}). The resulting expressions read
\bea
\label{eq:R1R2}
&& \hspace{-0.5cm}
R_1 = \dfrac{P_1}{r_0^{1/2}} \left[ 1 +  \dfrac{3}{4} \xi_{ll} \ln \xi_{ll} +
P_2 \xi_{ll} + \left(P_a + P_{as} \xi_{ss}\right) \dfrac{a^2}{r_0^2} + P_{ah}\,
a^2 m^2_{Ds} \right] +\dfrac{P_3}{r_0^{3/2} m_{D_s}}\,, \\
&& \hspace{-0.5cm} 
R_2 = P^{\prime}_1 \left[ 1 + \left(\dfrac{3}{4} (1+3 g_c^2) -
\dfrac{5}{4} \right) \xi_{ll} \ln \xi_{ll} + P^{\prime}_2 \xi_{ll} +
\left(P^{\prime}_a + P^{\prime}_{as} \xi_{ss}\right) \dfrac{a^2}{r_0^2} +
P^{\prime}_{ah}\, a^2 m^2_{Ds} \right] +\dfrac{P^{\prime}_3}{r_0
m_{D_s}}\,, \nn
\eea
where the coefficients $P_1$, $P_2$, $P^{\prime}_1$ and $P^{\prime}_2$ are then
expressed as linear functions of the strange quark mass as in
eq.~(\ref{eq:lin}).

The values of the coefficients $P_i$ and $P^{\prime}_i$ are collected in
Table~\ref{tab:parSU2HM} and the results for $f_D$, $f_{D_s}$ and $f_{D_s}/f_D$
are compared to those obtained from Fit I in Table~\ref{tab:resSU2HM}. The two
fits yield results that are in good agreement and with very similar
uncertainties.

\begin{figure}[t]
\vspace{-1.0cm}
\includegraphics[scale=0.28,angle=270,trim=75 0 0 0,
clip]{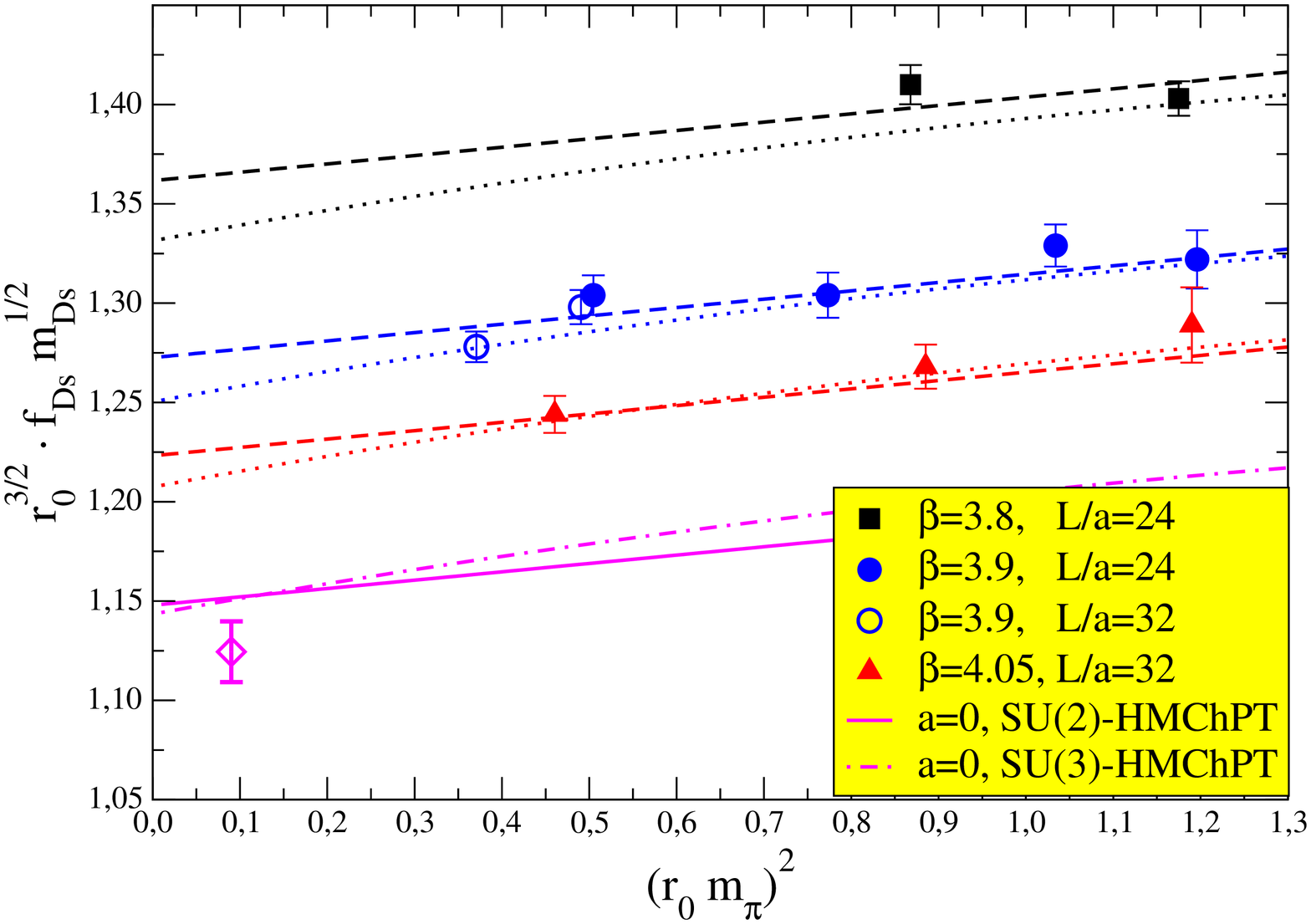}
\hspace{-0.8cm}
\includegraphics[scale=0.28,angle=270,trim=75 0 0 0, clip]{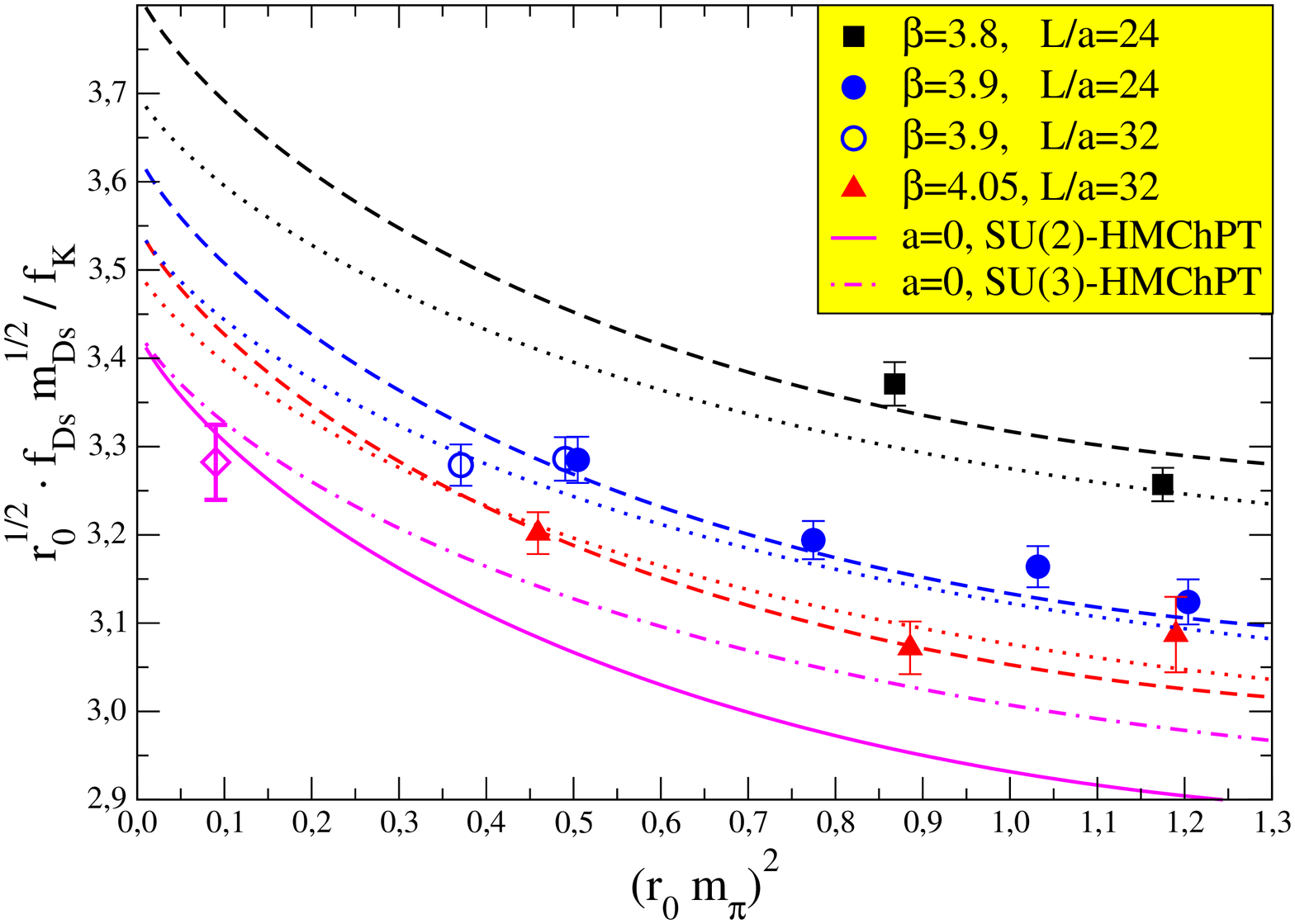} 
\vspace{-0.2cm} \\
\includegraphics[scale=0.28,angle=270,trim=75 0 0 0,
clip]{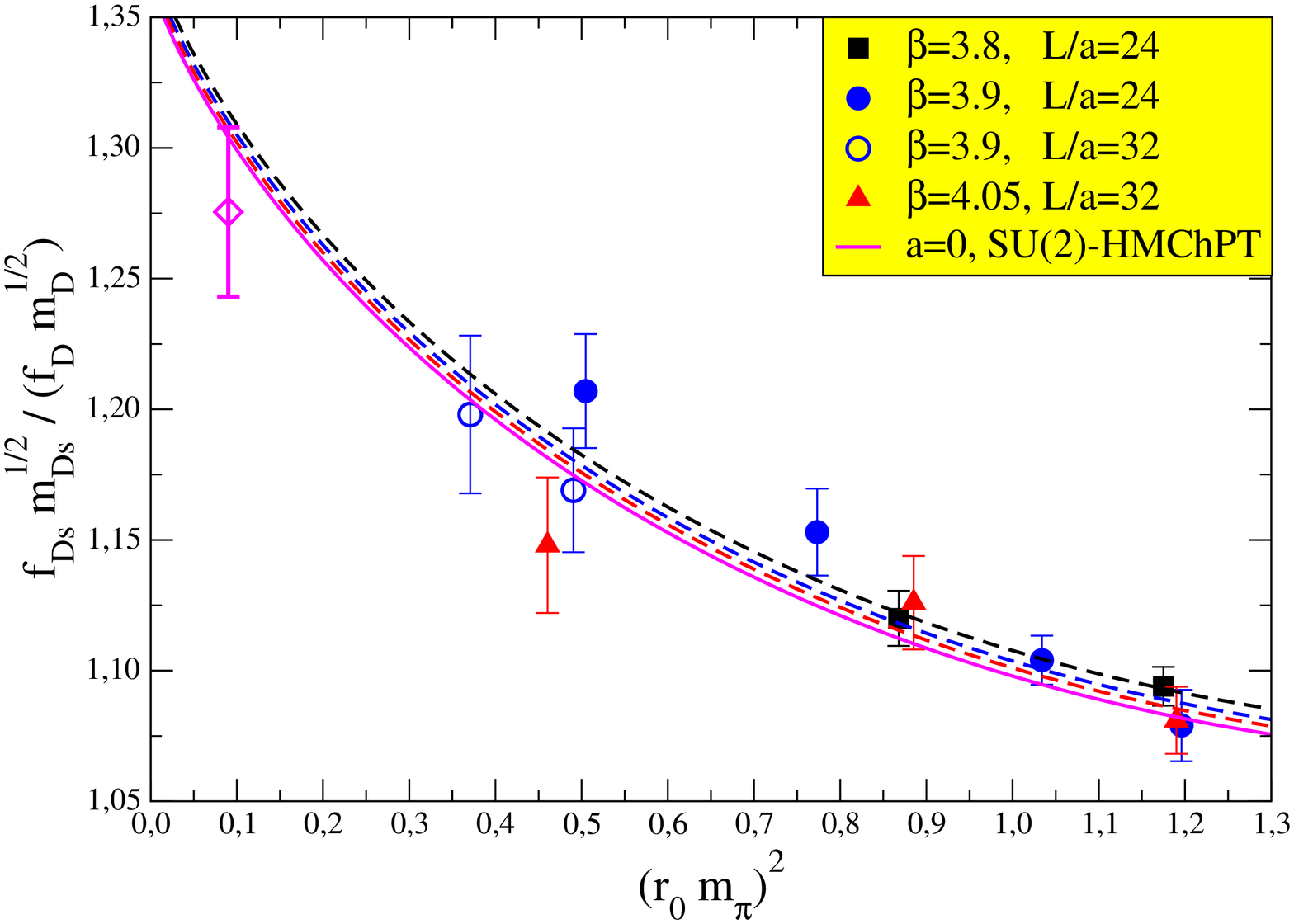}
\hspace{-0.8cm}
\includegraphics[scale=0.28,angle=270,trim=75 0 0 0, clip]{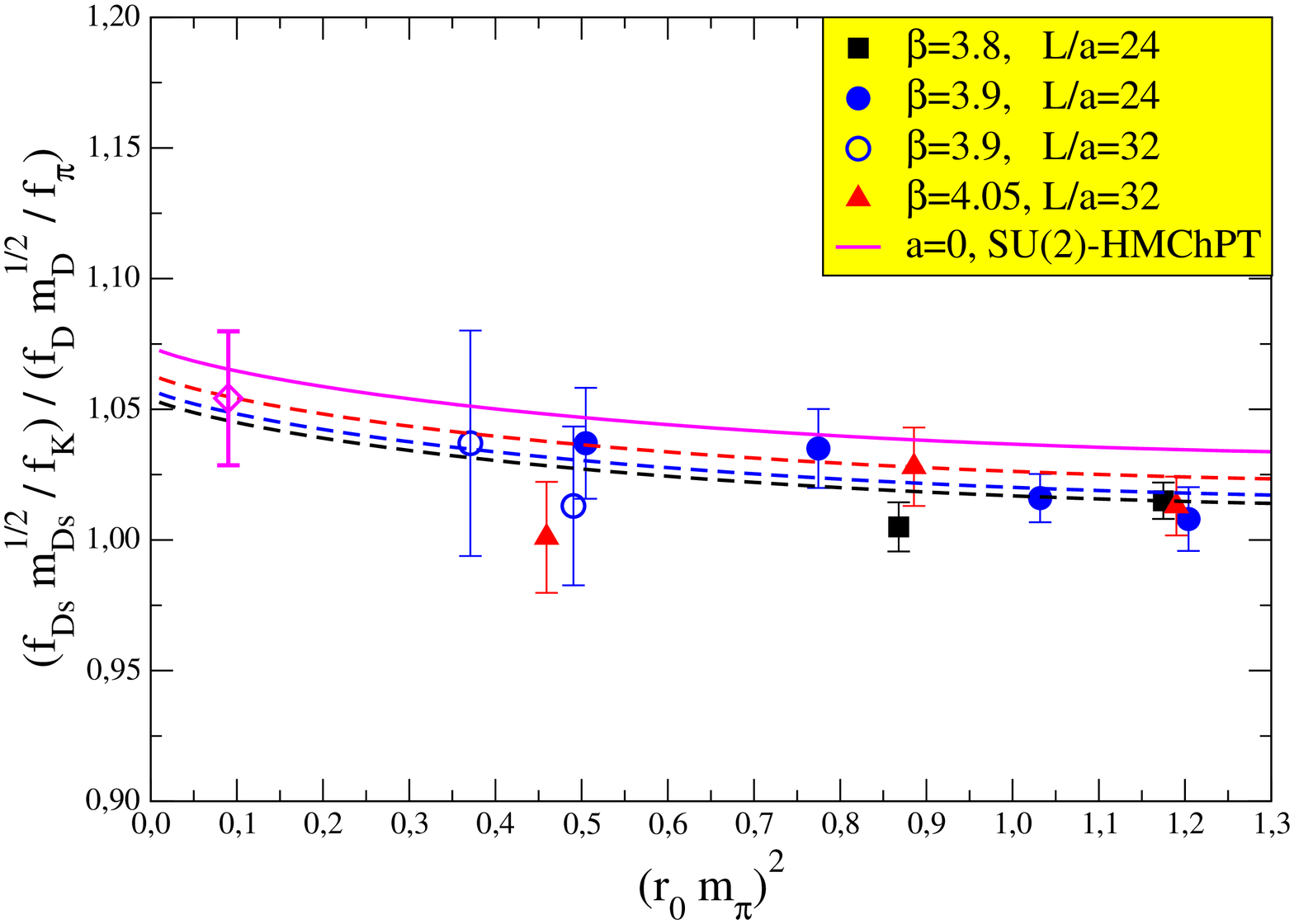}
\vspace{-0.4cm}
\caption{\sl From top-left to bottom-right: lattice results for $f_{D_s}
\sqrt{m_{D_s}}$, $R_1=f_{D_s} \sqrt{m_{D_s}}/f_K$, $R=f_{D_s}\sqrt{m_{D_s}}/(f_D
\sqrt{m_D})$ and $R_2=(f_{D_s} \sqrt{m_{D_s}}/f_K)/(f_D \sqrt{m_D}/f_{\pi})$ as
a function of the pion mass square $m_\pi^2 \equiv m_{PS}(\mu_l,\mu_l,\mu_l)^2$,
in units of $r_0$. We display data with $\mu_s$ and $\mu_c$ fixed to the
simulated masses that correspond to reference strange and charmed meson masses
$r_0 m_{PS}(\mu_l,\mu_s,\mu_s)=1.63$ and $r_0 m_{PS}(\mu_l,\mu_s,\mu_c)=4.41$.
The SU(2)- (SU(3)-) ChPT extrapolation to the physical pion mass is represented
at fixed lattice spacing by the dashed (dotted) curves, and in the continuum
limit by the solid (dashed-dotted) curve. The physical results, illustrated by
diamonds in the plots, are obtained from SU(2)-ChPT after interpolating to the
physical strange and charm quark masses.}
\label{fig:R_mll}
\end{figure}

In fig.~\ref{fig:R_mll} we show the dependence on the pion mass square of the
four quantities studied in Fits I and II.
We observe that $f_{D_s} \sqrt{m_{D_s}}$ (top-left) has a very mild dependence
on $m_{\pi}^2$, in agreement with the SU(2)-HMChPT prediction of
eq.~(\ref{eq:fDsSU2}) according to which chiral logarithms are absent for this
quantity at NLO. The logarithmic dependence in $f_{D_s} \sqrt{m_{D_s}}/f_K$
(top-right), thus comes only from the chiral logarithms predicted by SU(2)-ChPT
for the kaon decay constant, see eq.~(\ref{eq:fk}). The lattice results for the
double ratio $(f_{D_s} \sqrt{m_{D_s}}/f_K)/(f_D \sqrt{m_D}/f_{\pi})$
(bottom-right) are almost independent of the light quark mass. This is not
unexpected, since the chiral logarithms largely cancel in the ratio. We also
note that in the ratios $R$ and $R_2$ (bottom plots), where $f_{D_s}
\sqrt{m_{D_s}}$ is divided by $f_D \sqrt{m_D}$, discretization effects turn out
to be negligible, smaller than the statistical uncertainties.

As done for the light mesons decay constants, as an alternative to the
SU(2)-HMChPT approach, we have also tried to fit both the $D$ and the $D_s$
decay constants using SU(3)-HMChPT. The corresponding fitting formula is given
by eq.~(\ref{eq:fDsSU3}) with the addition of discretization terms. As for the
SU(2) case, we have tried two different fits, in which $f_D \sqrt{m_D}$ and
$f_{D_s} \sqrt{m_{D_s}}$ are either divided or not by the light decay constants
$f_\pi$ and $f_K$. At variance with our results for the light meson sector, we
find that the quality of the SU(3) fits, with $\chi^2/dof \simge 2$, is worse
than in the SU(2) case. For illustration, we show the results obtained from the
SU(3)-HMChPT analysis in the last line of Table~\ref{tab:resSU2HM} (for the case
in which the light decay constants are not introduced in the ratios) and in
fig.~\ref{fig:R_mll}. Even though these results are consistent with those
obtained from SU(2)-HMChPT, given the poor quality of the SU(3) fits, they are
not considered in deriving the final results.

\boldmath
\subsection{Results for $f_D$, $f_{D_s}$ and $f_{D_s}/f_D$}
\unboldmath
The results presented in Table~\ref{tab:resSU2HM} show that the SU(2)-HMChPT
analyses based on Fits I and II lead to determinations of $f_D$, $f_{D_s}$ and
$f_{D_s}/f_D$ that are in very good agreement, with very similar statistical
uncertainties. We choose to average these results and to quote their deviation
from the average as an additional systematic uncertainty due to the chiral
extrapolation.

As in the light meson case, we estimate the uncertainty due to residual
discretization effects by assigning an error of 50\% to the extrapolation from
our finest lattice at $\beta=4.05$ to the continuum limit. In the former case we
obtain $f_D^{\beta=4.05}=208\, \mev$ and $f_{D_s}^{\beta=4.05}=257\, \mev$, that
are $\simeq 5 \%$ above the continuum limit estimates. Note that this effect is
larger than the na\"ive estimate of leading discretization effects as being of
${\cal O}(\alpha_s a^2 \mu_c^2)$, which follows from the observation that ${\cal
O}((a \mu_c)^n)$ effects have been corrected at tree level in the definition of
the decay constants. This finding clearly illustrates the importance, for
lattice studies of heavy quarks, of evaluating discretization effects with
simulations performed at several values of the lattice spacing, rather than on
the basis of simple order of magnitude estimates. We also find that
discretization effects largely cancel in the ratio of the decay constants, and
we obtain $(f_{D_s}/f_D) ^{\beta =4.05} =1.23$ from Fit II, that is only $0.8\%$
below its continuum limit estimate. The same difference is even smaller in the
case of Fit I.

\begin{figure}[t]
\begin{center}
\vspace{-1.cm}
\includegraphics[scale=0.4,angle=270,trim=275 0 0 0, clip]{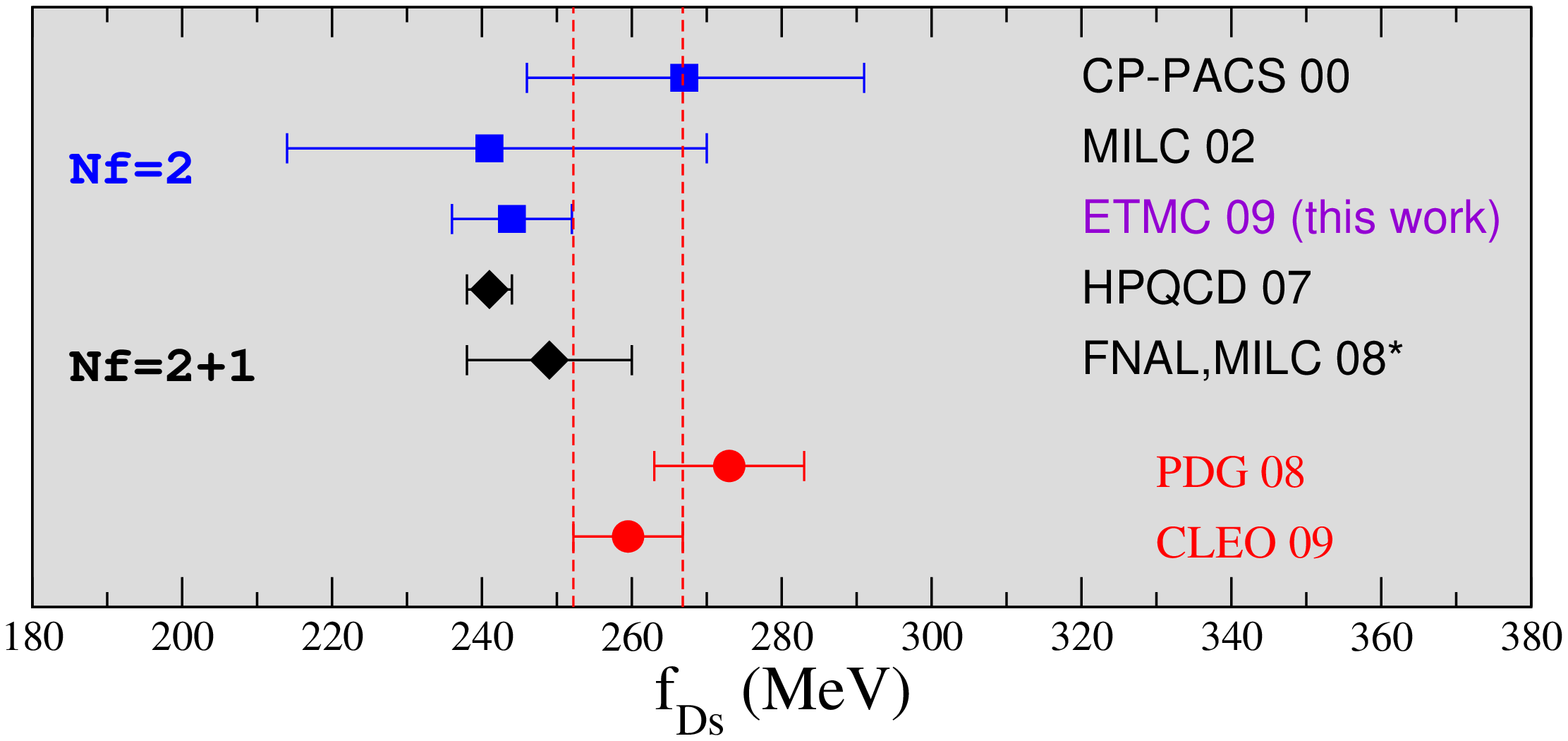} 
\includegraphics[scale=0.4,angle=270,trim=275 0 0 0, clip]{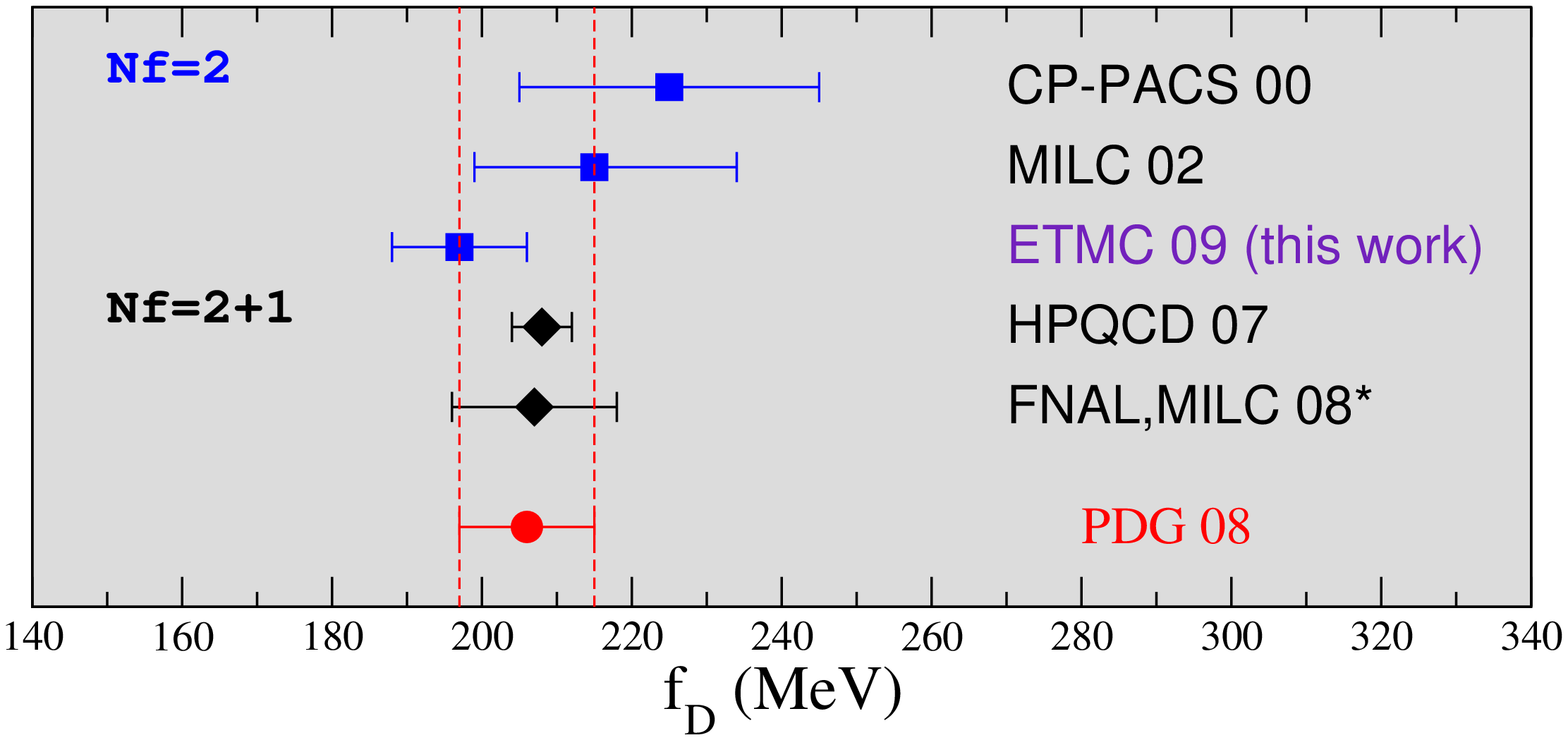}
\end{center}
\vspace{-1.0cm}
\caption{\small\sl Lattice QCD determinations of the $D$-mesons decay constants
$f_{D_s}$ (top) and $f_D$ (bottom) obtained from simulations with
$N_f=2$~\cite{AliKhan:2001jg,Bernard:2002pc} and
$N_f=2+1$~\cite{Follana:2007uv,Bernard:2009wr} dynamical fermions. A star in the
legend denotes preliminary results. The lattice results for $f_{D_s}$ are also
compared with the PDG 2008 experimental average~\cite{PDG} and with the recent
improved measurement by CLEO~\cite{Alexander:2009ux}. For $f_D$ we compare with
the CLEO determination~\cite{:2008sq}.}
\label{fig:fDunq}
\end{figure}
Our final results for the $D$ and $D_s$ decay constants and the ratio
$f_{D_s}/f_D$ are then
\beq
f_D= 197(6)(2)(6)\ \mev \quad , \quad f_{D_s}=244(3)(2)(7)\ \mev \quad , \quad
f_{D_s}/f_D=1.24(3)(0)(1) \,,
\eeq
where the errors come from statistics plus fitting, chiral extrapolation and
discretization effects, respectively. By combining all these uncertainties in
quadrature we finally obtain
\beq
f_D= 197(9)\ \mev \quad , \quad f_{D_s}=244(8)\ \mev \quad , \quad
f_{D_s}/f_D=1.24(3) \,.
\label{eq:fDfinal}
\eeq
The result obtained for $f_D$ is in very good agreement with the CLEO
measurement, $f_D^{exp.}=$ $205.8 (8.5) (2.5)\, \mev$~\cite{:2008sq}, and with
other $N_f=2$~\cite{AliKhan:2001jg,Bernard:2002pc} and
$N_f=2+1$~\cite{Follana:2007uv,Bernard:2009wr} lattice calculations, as shown in
fig.~\ref{fig:fDunq}. Even more interesting is the comparison shown in the same
figure between our result for $f_{D_s}$, other lattice results and the
experimental measurements. The PDG 2008 average was $f_{D_s}^{exp.}=273 (10)\,
\mev$~\cite{PDG}, higher than the values indicated by lattice calculations, for
which a possible explanation as an effect of new physics was given in
refs.~\cite{Dobrescu:2008er,Akeroyd:2009tn}. Recently, however, CLEO has
performed with higher statistics an improved measurement of the branching ratio
$Br(D_s^+ \to \tau^+ \nu \to e^+ \nu \bar \nu \nu)$~\cite{Onyisi:2009th} which,
combined with their measurements of $Br(D_s^+ \to \mu^+ \nu)$ and $Br(D_s^+ \to
\tau^+ \nu \to \pi^+ \bar \nu \nu)$, gives $f_{D_s}=259.5 (6.6)
(3.1)\,\mev$~\cite{Alexander:2009ux}. This latter determination, being in better
agreement with our and other lattice results, weakens the possibility of a new
physics effect in leptonic $D_s$ decays. In ref.~\cite{Alexander:2009ux}, also
an improved determination of the ratio of $D_s$ and $D$ decay constants is
provided by CLEO, $f_{D_s}/f_D=1.26 (6) (2)$, which is in good agreement with
our result in eq.~(\ref{eq:fDfinal}).

\section*{Acknowledgements}
We thank all the members of the ETM Collaboration for fruitful discussions. D.P.
thanks the Dipartimento di Fisica, Universit\`a di Roma Tre, and C.T. thanks the
Laboratoire de Physique Th\'eorique , Universit\'e de Paris XI, for the
hospitality. V.L., R.F., F.M. and S.S thank MIUR (Italy) for partial financial
support under the contract PRIN06. D.P. thanks MEC (Spain) for partial financial
support under grant FPA2005-00711. This work has been supported in part by the
EU Contract No.~MRTN-CT-2006-035482, ``FLAVIAnet'' and by the DFG
Sonderforschungsbereich/Transregio SFB/TR9-03. We also acknowledge  the
Consolider-Ingenio 2010 Program CPAN (CSD2007-00042).


\end{document}